\documentclass{aastex631}

\usepackage{mathrsfs,amssymb}
\usepackage{mathtools}
\usepackage{CJK}
\usepackage{amsmath}
\usepackage{graphicx}
\usepackage{natbib}
\usepackage{epsfig}

\usepackage{rotating}
\usepackage{longtable}

\newcommand {\Ks}  {{\rm K_S}}

\newcommand {\AK} {{A_{\K_S}}}

\newcommand {\lambdapeak}{\lambda_{\rm peak}}

\newcommand {\mum}{\,{\rm \mu m}}
\newcommand       \K            {\,{\rm K}}

\newcommand       \s            {\,{\rm s}}
\newcommand	\beq	{\begin{equation}}	
	\newcommand	\eeq	{\end{equation}}	

\newcommand       \simali       {\sim\,}

\submitjournal{ApJ}

\begin{document}
\begin{CJK*}{UTF8}{gbsn}

\title{Silicate Extinction Profile Based on the Stellar Spectrum by Spitzer/IRS}

\correspondingauthor{Biwei Jiang}
\email{bjiang@bnu.edu.cn}
\author[0000-0002-4448-3679]{Zhenzhen Shao (邵珍珍)}
\affiliation{Institute for Frontiers in Astronomy and Astrophysics,
            Beijing Normal University,  Beijing 102206, China}
\affiliation{Department of Astronomy,
               Beijing Normal University,
               Beijing 100875, China}
\affiliation{Department of Science Research,
                       Beijing Planetarium,
                       No. 138, Xizhimen Wai Street, Xicheng District,Beijing, 100044;
                       {\sf shaozhenzhen@bjp.org.cn}
                       }
\author[0000-0003-3168-2617]{Biwei Jiang (姜碧沩)}
\affiliation{Institute for Frontiers in Astronomy and Astrophysics,
            Beijing Normal University,  Beijing 102206, China}
\affiliation{Department of Astronomy,
               Beijing Normal University,
               Beijing 100875, China}



\begin{abstract}

The 9.7$\mum$ and 18$\mum$ interstellar spectral features,
arising from the Si--O stretching and O--Si--O bending mode of amorphous silicate dust,
are the strongest extinction feature in the infrared.
Here we use the "pair method" to determine the silicate extinction profile by comparing the \emph{Spitzer}/IRS spectra of 49 target stars with obvious extinction with that of un-reddened star of the same spectral type. The 9.7$\mum$ extinction profile is determined from all the 49 stars and the 18$\mum$ profile is determined from six stars. It is found that the profile has the peak wavelength around $\simali$9.2- 9.8$\mum$ and $\simali$ 18-22$\mum$ respectively. The peak wavelength of the 9.7$\mum$ feature seems to become shorter from the stars of late spectral type, meanwhile the FWHM seems irrelevant to the spectral type, which may be related to circumstellar silicate emission. The silicate optical depth at 9.7$\mum$, $\Delta\tau_{9.7}$,  mostly increases with the color excess in $J-K_S$ ($E_{\rm JK_S}$). The mean ratio of the visual extinction to the 9.7$\mum$ silicate absorption optical depth is $A_{\rm V}/\Delta\tau_{9.7}\approx17.8$, in close agreement with that of the solar neighborhood diffuse ISM. When $E_{\rm JK_S} \textgreater$4, this proportionality changes. The correlation coefficient between the peak wavelength and FWHM of the 9.7$\mum$ feature is 0.4, which indicates a positive correlation considering the uncertainties of the parameters. The method is compared with replacing the reference star by an atmospheric model SED and no significant difference is present.
%

\end{abstract}



\section{Introduction} \label{sec:intro}

%
Silicate as well as carbon dust is the major component of interstellar dust. It can be generally divided into two types, amorphous and crystalline by their structure, while the highly dominant type in the ISM is amorphous \citep{li2001} that shows two prominent broad absorption features around about 9.7 and 18 $\mum$ respectively (e.g. \citealp{gillett1975,kemper2004,chiar2006,min2007}) due to the Si-O stretching and O-Si-O bending vibration. In comparison, crystalline silicate exhibits multiple distinct narrow bands from $\sim 10-60 \mum$ \citep{molster2005,henning2010,liu2017} in almost all circumstellar environments.

The extinction profile caused by interstellar silicate has been analysed with different samples, which lead to inconsistent results on the peak wavelength $\lambda_{\rm peak}$ and the FWHM.

\citet{kemper2004} analyzed the $\simali$8-13$\mum$ spectrum of the Galactic center source Sgr A$^\star$ obtained with the Short Wavelength Spectrometer (SWS) on board the Infrared Space Observatory (ISO). They derived the 9.7$\mum$ silicate absorption profile of Sgr A$^\star$ with the peak at $\simali$9.8 $\mum$ and the FWHM of $\simali$1.73 $\mum$ by subtracting a fourth-order polynomial continuum from the observed spectrum. They suggested that the narrow profile could be caused by the contamination of the silicate emission intrinsic to the Sgr A$^\star$ region. \citet{chiar2006} studied the $\simali$2.38-40$\mum$ ISO/SWS spectra of the diffuse ISM along the lines of sight towards four heavily extinguished WC-type Wolf-Rayet stars. They found that the 9.7$\mum$ silicate absorption features of the four sources all peak at $\simali9.8\mum$, but their widths vary from $\simali2.35\mum$ for WR98a to $\simali2.7\mum$ for WR104, apparently wider than $\simali1.73\mum$ of Sgr A$^\star$.

\citet{mcclure2009} derived the $\simali$5-20$\mum$ extinction curves from 28 G0-M4 III stars lying behind the Taurus, Chameleon I, Serpens, Barnard 59, Barnard 68, and IC5146 molecular clouds by comparing the observed spectrum by the Infrared Spectrograph (IRS) on board the Spitzer Space Telescope with the stellar photospheric model spectrum of \citet{castelli1997}. She found that the silicate extinction profile in the regions of $\AK \textless$  1mag peaks at $\simali$9.63 $\mum$ and has a FWHM of $\simali$2.15 $\mum$ whereas, in regions of 1  $\textless \AK \textless$  7mag, it peaks at $\simali$9.82 $\mum$ and has a FWHM of $\simali$2.72 $\mum$. This means that the 9.7$\mum$ silicate extinction feature appears to be broadened in more heavily extinguished regions.
\citet{olofsson2011} derived the mid-IR extinction curve for a highly obscured M giant (\# 947) behind the dark globule B335 ($R_V$ $\approx$ 4.9, $A_{\rm V}$ $\simali$10 mag), using the $\simali$7-14$\mum$ Spitzer/IRS spectrum complemented by the $\simali$5-16$\mum$ spectrum obtained with the ISOCAM/CVF instrument on board ISO. Their result is even more special in that the 9.7$\mum$ feature peaks at $\simali$9.2 $\mum$ and has a FWHM of $\simali$3.80 $\mum$, i.e. the ever shortest peak wavelength and the largest width of the silicate profile.
Also based on the Spitzer/IRS data, \citet{vanbreemen2011} investigated the silicate absorption spectra of three sightlines towards diffuse clouds and four sightlines specifically towards the Serpens, Taurus, and ρ Ophiuchi molecular clouds. They found that the 9.7$\mum$ silicate absorption bands in the diffuse sightlines show a strikingly similar band shape and all closely resemble that of Sgr A$\ast$  \citep{kemper2004}, while the 9.7$\mum$ features in the molecular cloud sightlines differ considerably from that of Sgr A$\ast$ by peaking at $\simali$9.72 $\mum$ and having a FWHM of $\simali$2.4 $\mum$. More recently, \citet{fogerty2016} analyzed the Spitzer/IRS spectra of the 9.7$\mum$ silicate optical depths of the diffuse ISM along the line of sight towards Cyg OB2-12, a heavily extinguished luminous B5 hypergiant with $A_{\rm V}$ $\approx$ 10.2 mag, and towards $\zeta$ Ophiuchi, a lightly extinguished bright O9.5 star with $A_{\rm V}$ $\approx$ 1mag. They found appreciable differences between the spectral profile of the 9.7$\mum$ silicate absorption towards Cyg OB2-12 and $\zeta$ Ophiuchi; while the former peaks at $\simali$9.74 $\mum$ and has a FWHM of $\simali$2.28$\mum$, the latter peaks at $\simali$9.64 $\mum$ and has a FWHM of $\simali$2.34 $\mum$. Moreover, the contrast between the feature and the absorption continuum of the former exceeds that of the latter by $\simali$30 per cent. Recently \citet{hensley2020} show a new analysis of archival ISO-SWS and Spitzer IRS observations of Cyg OB2-12 using a model of the emission from the star and its stellar wind to determine the total extinction $A_\lambda$ from 2.4-37 $\mum$. They derived the FWHM of the 9.7$\mum$ silicate absorption profile being 2.23 $\mum$, consistent with the result of \citet{fogerty2016}.

We investigated the 9.7$\mum$ silicate absorption profile using the Spitzer/IRS observations of five early-type stars 
and found that their peak wavelength concentrates at 9.7$\mum$, and the FWHM is divided into two groups, 2 $\mum$ for three stars and 3$\mum$ for two stars \citep{shao2018}. The possible reasons for the diversity were discussed but no solid conclusions could be drawn.

In summary, previous studies about the peak and FWHM of silicate dust extinction profile yield inconsistent conclusions. The peak varies from   $\simali$9.2 to  $\simali$9.82$\mum$, and the range of FWHM fluctuates even more from $\simali$ 1.73 to  $\simali$ 3.8$\mum$. Moreover, no pattern of variation could be found. From the technical side, various works use different data, including the dispersion of stellar type, sightline direction, and correspondingly interstellar environment. In addition, the method to derive the extinction profile is different. In this work, we try to investigate the systematic variation of the silicate extinction profile by including a significantly large sample of tracing stars which span a wide range of spectral type and sightlines.
%
%
The method will be described in Section 2,  and in Section 3 the data is introduced. Then the results and discussion will be shown in Section 4.

\section{Method} \label{sec:method}

The interstellar UV/optical extinction curve is often determined by comparing the spectrum
of a reddened star with that of an unreddened star of the same spectral type, which is the so-called "pair method". There are additional methods to derive the intrinsic spectrum, including the atmosphere model and polynomial fitting \citep{mcclure2009,kemper2004}. In this work, we continue to use the “pair method” to calculate the ratio of the color excess $E(\lambda-K_S)/E(J-K_S)$ where $J$ and $K_S$ are the two near-infrared bands centering at 1.2 and 2.2$\mu$m respectively. The details of this method were described by \citet{shao2018}. In brief, the spectrum of the target source that has obvious silicate extinction is compared with a reference source that is of the same spectral type and the same luminosity class but with negligible extinction. However, such reference star cannot always be found because the Spitzer/IRS spectrum is available only for a limited number of stars. In such case, spectral type has the priority over luminosity class since it is the primary factor to decide the intrinsic spectral energy distribution.  The one-to-one pair is listed in Table 1 with their spectral type, which shows that most of the targets are well paired with the reference.

The apparent stellar spectrum $F_\lambda$ is the intrinsic spectrum $F^0_{\lambda}$ dimmed by the interstellar extinction $A_\lambda$ and the geometrical distance $d$:
\begin{equation}
	F_\lambda = \frac{F^0_{\lambda}\cdot
		\exp\left(-A_\lambda/1.086\right)
		\cdot \pi R^2}
	{4\pi d^2}~~,
\end{equation}
where $R$ is stellar radius.
If the reference star has the same intrinsic spectrum as the target star, then the intrinsic flux ratio of the star in the $K_S$ and $\lambda$ band can be replaced by that of the reference star, i.e.,
$F^0_{K_S}/F^0_{\lambda} = F^{\rm ref}_{K_S}/F^{\rm ref}_\lambda$, where ``ref'' refers to the reference star.
By comparing the observed spectrum of the source star with the reference star, the color excess between the $\lambda$ and $K_S$ band, $E(\lambda-K_S)$, is obtained as:
\begin{equation}\label{eq:EKslambda}
	\begin{aligned}
		E(\lambda-K_S) =  A_\lambda-A_{K_S}
		= -2.5\log\left(\frac{F_\lambda}{F_{K_S}} \frac{F^{\rm ref}_{K_S}}{F^{\rm ref}_\lambda}\right)\\
		= -2.5\log\left(\frac{F_\lambda}{F^{\rm ref}_{\lambda}}\right)
		+ 2.5\log\left(\frac{F_{K_S}}{F^{\rm ref}_{K_S}}\right) \\
		= -2.5\log\left(\frac{F_\lambda}{F^{\rm ref}_{\lambda}}\right)
		+ m^{\rm ref}_{K_S} - m_{K_S} ~~,
	\end{aligned}
\end{equation}
where $m_{K_S}$ and $m^{\rm ref}_{K_S}$ are respectively the apparent $K_S$-band magnitudes of the source star and the reference star.
Since we are considering the color excess $E(\lambda-K_S)$ instead of the absolute extinction $A_\lambda$, the distance $d$ to the source star (or the reference star) is cancelled out in Eq.\,\ref{eq:EKslambda}.

After the extinction curve is determined, the peak wavelength and the FWHM of the $9.7\mum$ and $18\mum$ features are calculated in the same way as \citet{shao2018}, i.e. by subtracting the continuum. The continuum is fitted by a power law to the brightness in the $JHK_S$ bands and 6-7.5$\mum$ extinction with an index of -1.06 \citep{gordon2023one}. It should be mentioned that the method to determine the continuum would influence the results. Linear fitting has also been adopted in other works, which works well for high quality spectrum. A constant index implies that the continuum extinction law does not change with the source, which may not be true. However, as we are interested in the silicate profile, this can reflect its relative variation.

\section{The targets} \label{sec:data}

\begin{figure*}
	\vspace{-1mm}
	\centering
	\includegraphics[width={17.2cm}]{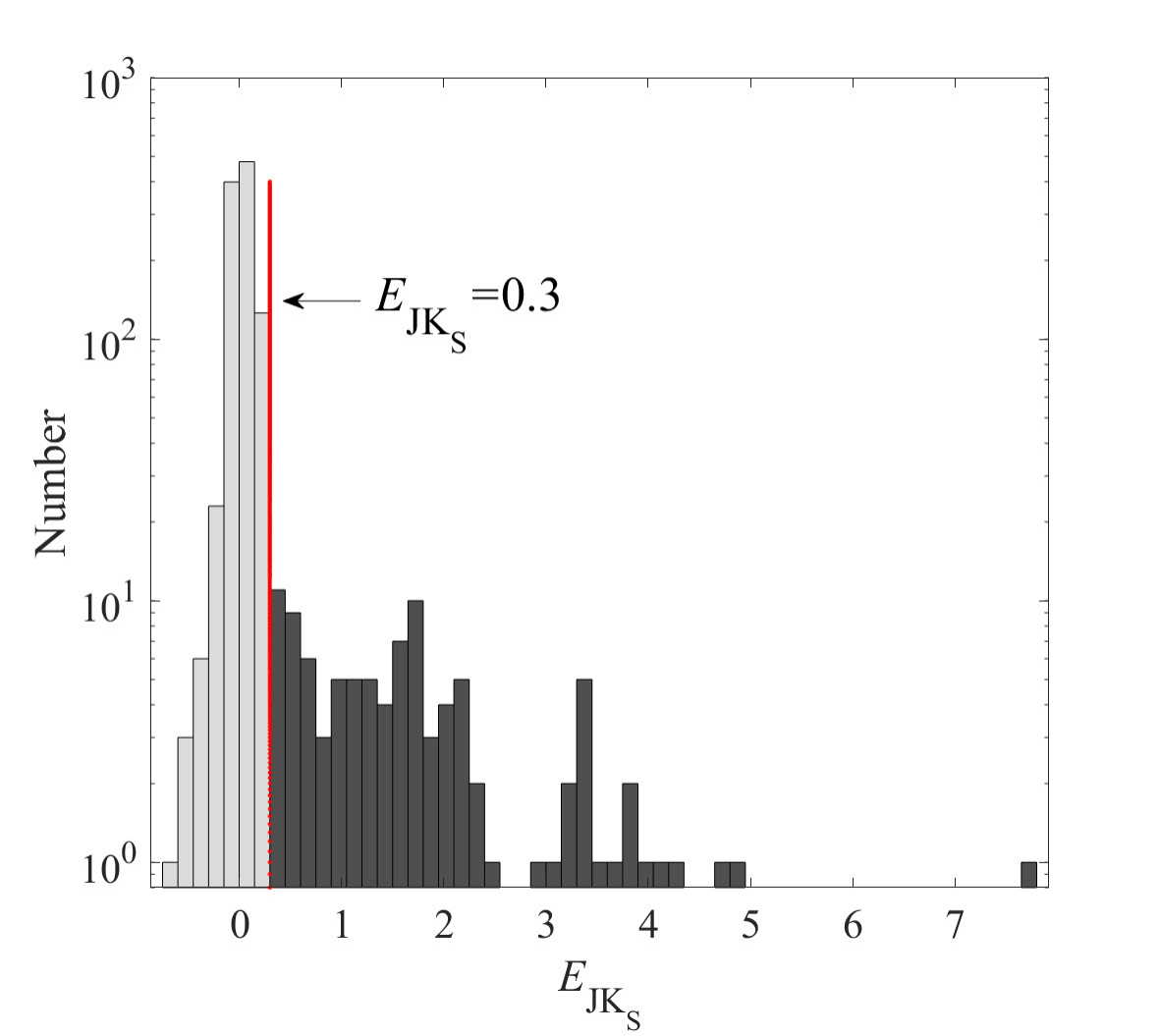}
	
	\caption{
		\label{fig:EJK}
		The histogram of $E_{\rm JK_S}$ of the stars in the initial sample. The $E_{\rm JK_S} = 0.3$ is labeled by the red line for selecting the targets with visible silicate absorption feature.
	}
	\vspace{4mm}
\end{figure*}

The data are taken from the $\sim$16,900 low-resolution  Spitzer/IRS spectra \citep{houck2004,werner2004spitzer}. These spectra were merged from four slits: SL2 ($\sim$5.21-7.56$\mum$), SL1 ($\sim$7.57-14.28 $\mum$), LL2 ($\sim$14.29-20.66$\mum$), and LL1 ($\sim$20.67-38.00$\mum$), and some spectra are available only in the first two or three slits. \citet{chen2016} cross-identified the types of these objects in the SIMBAD database, supplemented with the photometry by the 2MASS and WISE all-sky surveys. This cross-identification resulted in a database of 126 O stars, 414 B stars, 806 A stars, 453 F stars, 543 G stars, 1397 K stars, and 1260 M stars, which forms the initial catalog of our selection of target and reference stars. We take the following approaches to select our sample:
\begin{enumerate}
  \item Exclude the sources showing silicate and/or polycyclic aromatic hydrocarbon (PAH) emission features, which indicates the presence of circumstellar dust. This reduces the sample to be consisted of 55 O stars, 203 B stars, 806 A stars, 453 F stars, 543 G stars, 1397 K stars, and 1260 M stars.
  \item Require the signal-to-noise ratio (S/N) of the Spitzer/IRS spectrum in SL1 and SL2 to be $\geq25$. The signal-to-noise ratio is quoted from the IRS spectrum, and for more specific information see \citep{houck2004}. For multiple observations, we select the observation with the highest SNR and remove other repeated observations. Stacking the repeated observation may increase the S/N, but taking one single spectrum can have the highest accuracy because different observations can have systematic error due to, e.g. instrument instability and calibration process. Indeed, the three sources with repeated observations, specifically Elia 3-3, BD+43 3770 and HD 147889, all had the highest SNR $>$ 70, and no stacking process is performed. Nevertheless, both stacking and single spectrum is used by various works. This removes 5 O stars, 11 B stars, 47 A stars, 60 F stars, 97 G stars, 209 K stars, and 51 M stars.
  \item Require the color excess $E(J-K_S)>0.3$ mag, corresponding to a line-of-sight extinction of $\AK \textgreater $ 0.2 mag, which is comparable to three times of the photometric error.  While the peak extinction of the 9.7$\mum$ silicate feature is comparable to $\AK$, the extinction of the blue and red wings of the 9.7$\mum$ silicate feature drops by a factor of $\textgreater $2 (e.g., see \citealp{xue2016}). With $\AK \textgreater $  0.2mag, the entire 9.7$\mum$ silicate extinction profile should be measurable. The color excess $E_{\rm JK_S}$ (the short for $E(J-K_S))$ is calculated by using the intrinsic color index $C^0_{\rm JK_S}$ according to the object's spectral type listed in the Allen's Astrophysical Quantities \citep{allen2000}. Figure \ref{fig:EJK} displays the distribution of $E_{\rm JK_S}$, and it can be seen that only a third of the sources satisfy this criterion. The $E_{\rm JK_S}$ of some sources is less than 0, which may be caused by the photometric error as most negative $E_{\rm JK_S}$ is within the range of three times the photometric error. The median of the negative $E_{\rm JK_S}$ is -0.04, which can be taken as the uncertainty of this color excess. This uncertainty would be transferred to 0.02 mag in $\AK \sim 0.5E_{\rm JK_S}$. This criterion $E(J-K_S)>0.3$ corresponds to a significance level of 7 sigma. This restriction removes more sources, leaving only 13 O-type stars, 21 B-type stars, 4 A-type stars, 5 F-type stars, 4 G-type stars, 12 K-type stars, and 44 M-type stars.
\end{enumerate}

\begin{figure*}
	\vspace{-1mm}
	\centering
	\includegraphics[width={16.2cm}]{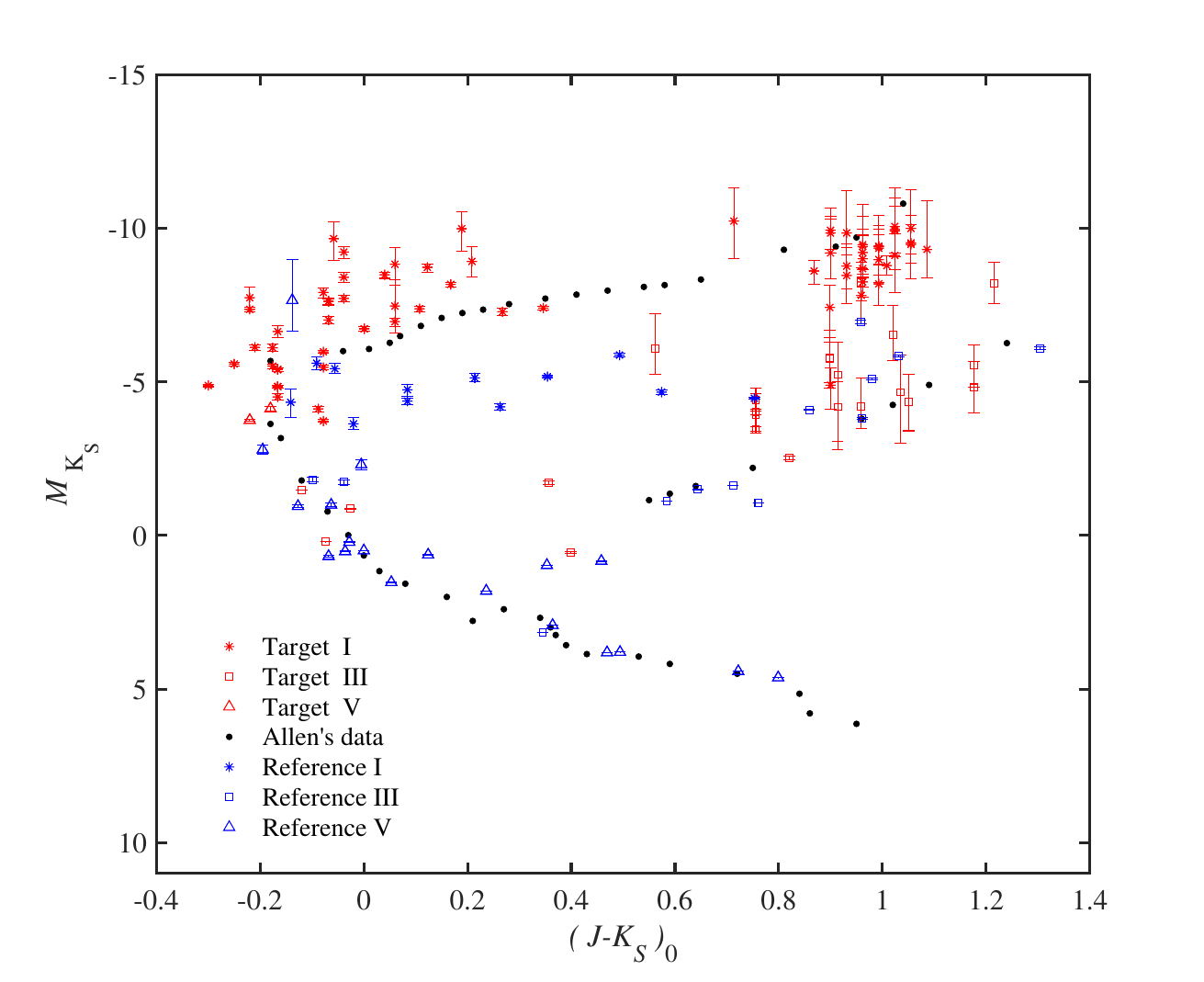}
	
	\caption{
		\label{fig:HRD}
		The color-magnitude diagram of the target and reference stars. The black dots are taken from Allen's Astrophysical Quantities for various types of stars that represents the standard track, and the red and blue dots are for the target and reference stars respectively. The luminosity class I, III and V are denoted by asterisk, diamond and triangle respectively.
	}
	\vspace{4mm}
\end{figure*}

\begin{figure*}
	\vspace{-1mm}
	\centering
	\includegraphics[width={16.2cm}]{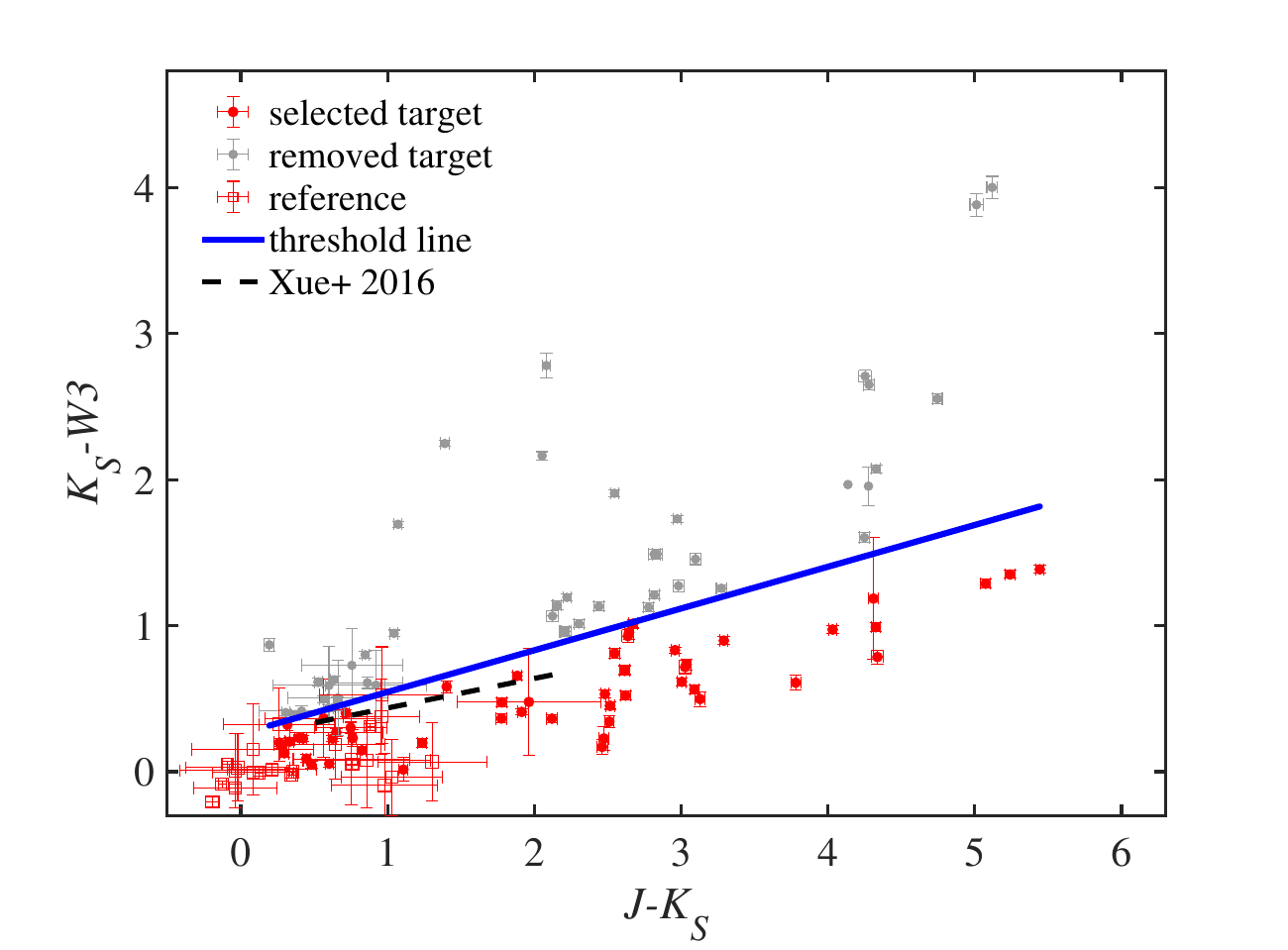}
	
	\caption{
		\label{fig:kw3jk}
		The two color diagram of $K_S-W3$ vs $J-K_S$ of the preliminary sample target stars and the reference stars. The stars above the solid line may have circumstellar silicate emission and are removed.  The dashed-line is taken from \citet{xue2016}.
	}
	\vspace{4mm}
\end{figure*}

In addition, silicate dust exists in circumstellar envelop as well as in interstellar medium. The presence of circumstellar silicate can deform interstellar extinction mostly by emission. With the geometric distance from the Gaia/EDR3 catalog and correcting the extinction $\AK$ converted from $E_{\rm JK_S}$, the color-magnitude diagram, $J-K_S$ vs. $M_{\rm K_S}$ of the above selected stars is displayed in Figure \ref{fig:HRD}. Apparently, some stars are red (super)giants that may have circumstellar dust resulted from stellar wind, which should be removed. \citet{xue2016} found that the stars with circumstellar silicate emission appear to display apparent color excess at $K_S-W3$ ($W3$ is the WISE band centering around 12$\mum$) in the $J-K_S$ vs. $K_S-W3$ diagram (see Fig. 19 in \citealp{xue2016}) due to the silicate broad features around 9.7$\mum$. Following this criterion, the $K_S-W3$ vs. $J-K_S$  diagram is plotted in Figure \ref{fig:kw3jk} to identify the emission of the circumstellar silicate. In general, the color index $K_S-W3$ is proportional to $J-K_S$ as expected from interstellar extinction. However, the existence of circumstellar silicate emission would increase the brightness in $W3$ and correspondingly the color index $K_S-W3$. According to the reference stars, the threshold line is determined by connecting the two points with the largest color index $K_S-W3$, which also coincides with the trend line of \citet{xue2016} to determine whether there is circumstellar dust emission (see  Figure \ref{fig:kw3jk}). The stars above the line are considered to have circumstellar dust emission and removed. It should be kept in mind that this operation cannot remove all the sources with silicate emission since an interstellar absorption can weaken the emission and lead to a normal color index $K_S-W3$.

\begin{figure*}
	\vspace{-1mm}
	\centering
	\includegraphics[width={8cm}]{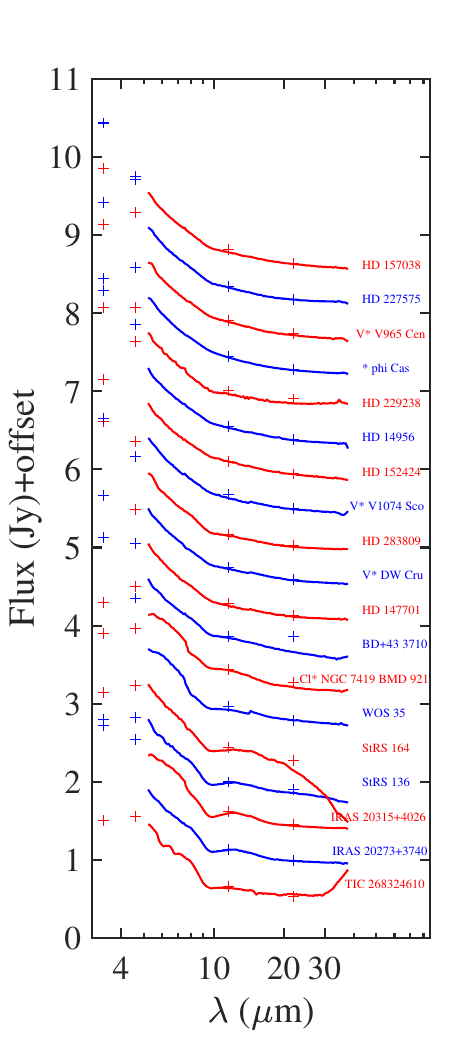}
	\includegraphics[width={8cm}]{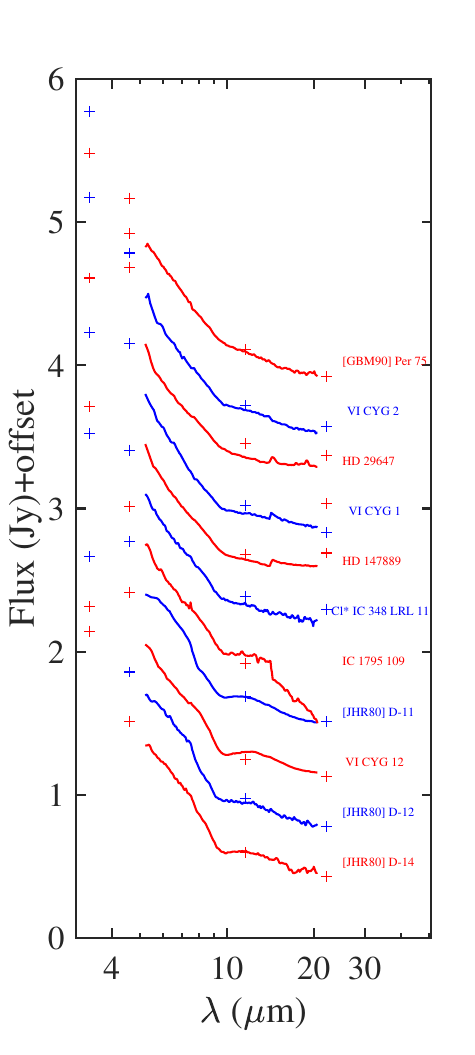}
	
	\caption{
		\label{fig:flux40}
		The Spitzer/IRS spectrum that ends at 38$\mum$ and 21$\mum$ together with the WISE photometric brightness.
	}
	\vspace{4mm}
\end{figure*}

\begin{figure*}
	\vspace{-1mm}
	\centering
	\includegraphics[width={8cm}]{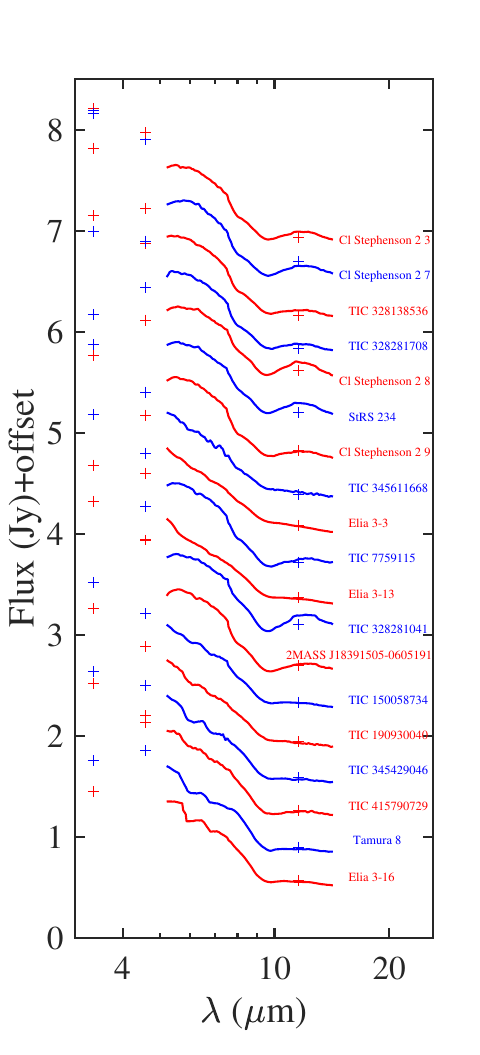}
	\includegraphics[width={8cm}]{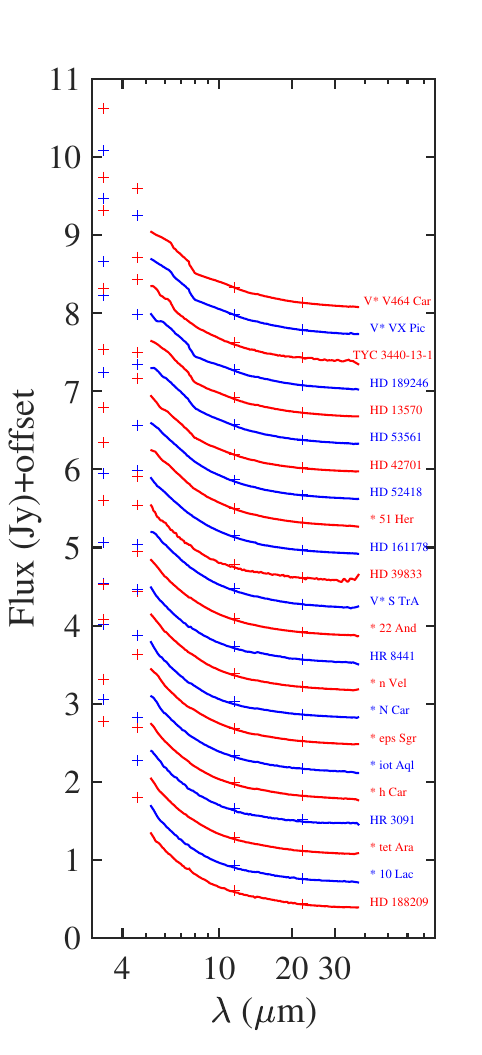}
	
	\caption{
		\label{fig:flux14}
		The same as Figure \ref{fig:flux40} except that the spectrum on the left ends at 14$\mum$ and the reference stars spectrum on the right.
	}
	\vspace{4mm}
\end{figure*}

Finally, we select 49 stars as the target sources for studying silicate dust extinction, consisted of five O stars, thirteen B stars, three F stars, four G stars, nine K stars, and fifteen M stars. Their infrared photometric magnitudes are listed in Table \ref{table:t2} together with the reference stars. Their spectrum flux reliability is further examined by the WISE photometry (c.f Table 2). For each target star, the photometric fluxes in the four WISE bands at 3.35, 4.60, 11.56, and 22.08 $\mum$ agree with the Spitzer/IRS spectra for most stars. A few stars' photometric fluxes in the W4 band deviate from the spectra  as shown in Figure \ref{fig:flux40} and \ref{fig:flux14} for those ending at 38$\mum$, 21$\mum$ and 14$\mum$ separately,  which is attributed to the photometric error in the W4 band due to its relatively poor quality.

\section{Result \& Discussion}

\subsection{Characteristics of the Silicate Extinction Profile}

\subsubsection{The Profile around $9.7\mum$ }

\begin{figure*}
	\vspace{-1mm}
	\centering
	\includegraphics[width={17.0cm}]{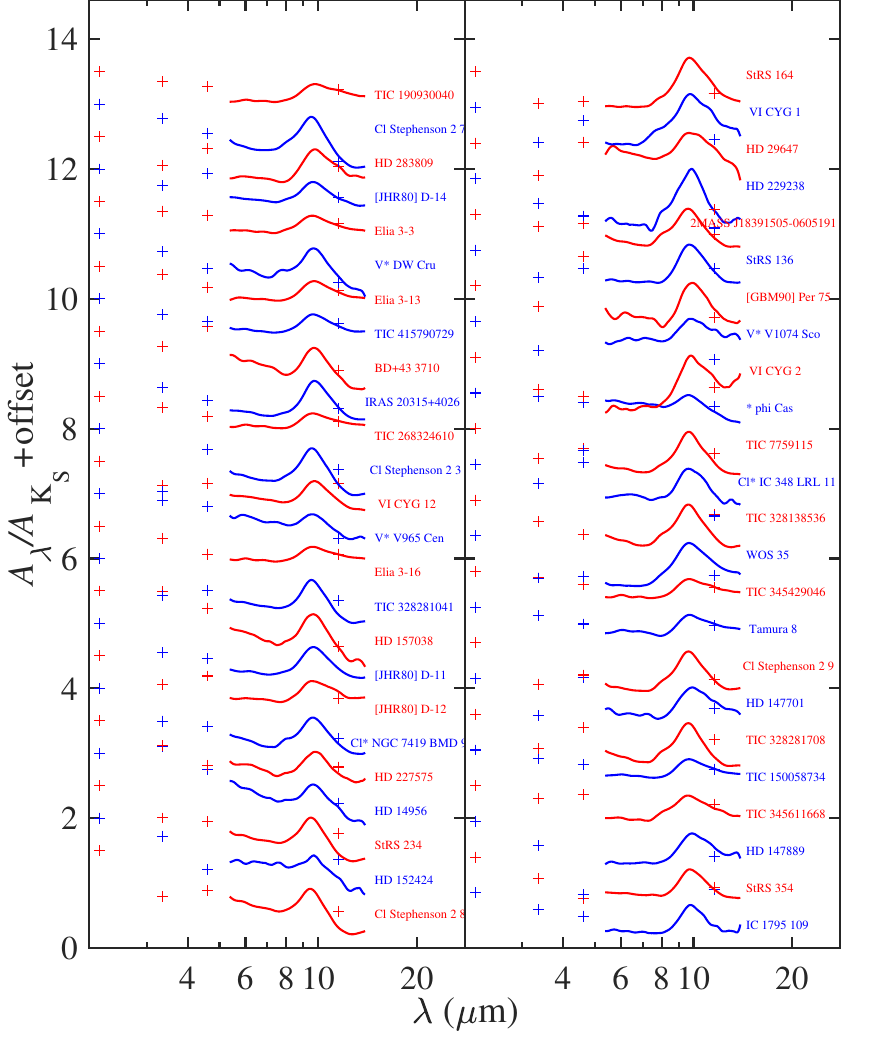}
	
	\caption{
		\label{fig:extincurve12}
		The extinction curves that only include the 9.7$\mum$ profile from 49 stars. The offset here is standardized on the K-band, moving equally spaced by 0.5.
	}
	\vspace{4mm}
\end{figure*}

\begin{figure*}
	\vspace{-1mm}
	\centering
	\includegraphics[width={11cm}]{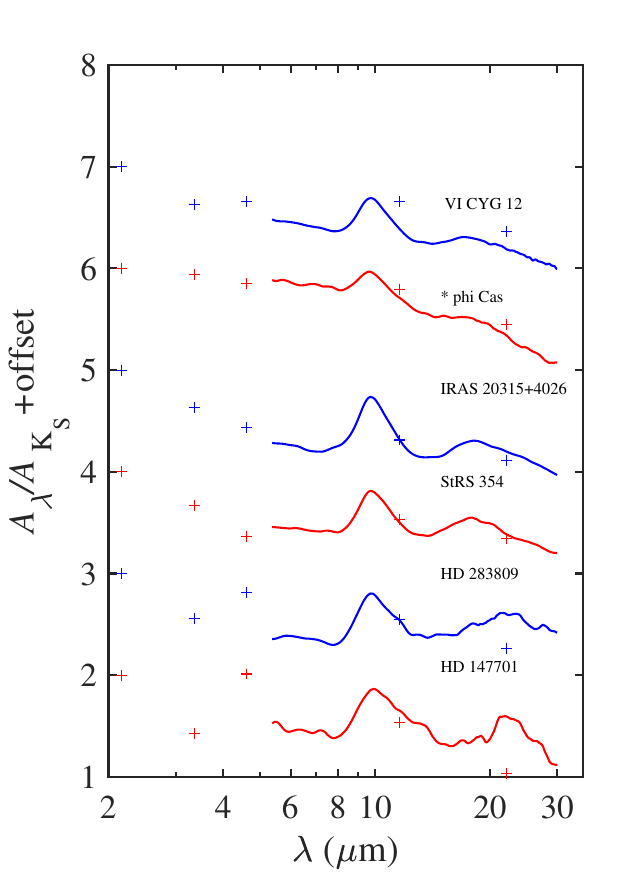}
	
	\caption{
		\label{fig:extincurve18}
			The extinction curves that contain both the 9.7$\mum$ and 18$\mum$ profiles from  six stars. The offset here is standardized on the K-band, moving equally spaced by 1.
	}
	\vspace{4mm}
\end{figure*}

The derived 49 extinction curves are displayed in Figure \ref{fig:extincurve12} for the $9.7\mum$ feature only and Figure \ref{fig:extincurve18} for those including the $18\mum$ feature.  All the 49 curves cover the $9.7\mum$ feature with reliable results, meanwhile only six of them show a relatively evident 18$\mum$ profile because of the much smaller extinction and the limitation of data quality at longer wavelength.
The peak wavelength ($\lambdapeak$), full width at half maximum ($FWHM$), and optical depth at $9.7\mum$ ($\Delta\tau_{9.7}$) of the silicate feature are listed in Table 3. It is worth noting that the optical depth $\Delta\tau_{9.7}$ and FWHM here are those of the silicate features after subtraction of the continuum spectrum and the optical depths $\Delta\tau_{9.7}$ refers to the value at the actual peak. The parameter definitions for the 18$\mum$ feature are alike.

\begin{figure*}
	\vspace{-1mm}
	\centering
	\includegraphics[width={19.0cm}]{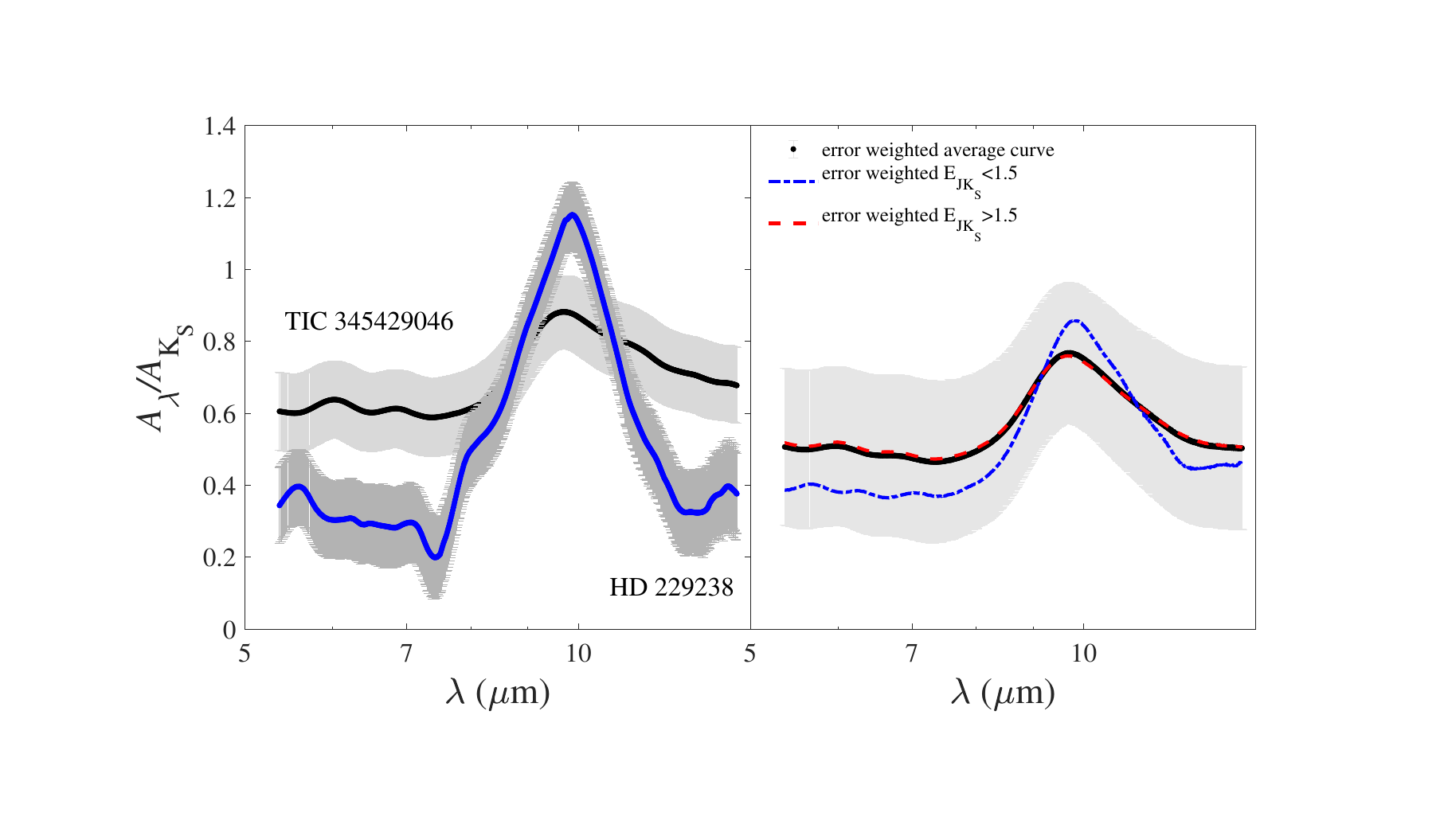}
	
	\caption{
		\label{fig:extincurveerror}
	Left: Two examples of the extinction curve with errors. Right: This figure is the comparison of different mean extinction curve calculation. The black extinction curve is error weighted average curve by $1/\sigma^2$ with error bars. The shaded area represents the error of the weighted leveling curve at 1 sigma, where the sigma is obtained as the median sigma of the corresponding 49 sources at the same wavelength. The blue dotted line is the weighted average extinction curve when $E(J-K_S)$ is less than 1.5, and the red dotted line is the weighted average extinction curve when $E(J-K_S)$ is greater than 1.5.}
	\vspace{4mm}
\end{figure*}

The uncertainties of these feature parameters are calculated by the bootstrap method which repeats the analysis 1000 times by re-sampling, i.e. by adding synthetic noise to the previously-taken observations and reanalyzing. The synthetic noise follows the prior that the noise amplitudes obey the Gaussian distribution around the observed IRS flux with the sigma being the flux error. The standard deviation of the 1000 results is taken as the uncertainty ($\sigma$ later) of the curve. The left panel of Figure \ref{fig:extincurveerror} shows two extinction curves with the error of TIC 345429046 and HD 229238. Depending on the spectral and photometric data quality of the target and reference star, the error of the extinction curve differs significantly. The error for HD 229238 is relatively small, on the order of 0.05 in $A_\lambda/\AK$.  However, for the case that both the target and reference stars have large error, the error of  $A_\lambda/\AK$ can be up to 0.4. The two examples in left panel also show that the extinction curves of different line of sight directions are different, although taking into account the existence of errors, the actual difference may be smaller than it looks, but this difference cannot be fully explained by the error. The right panel of Figure \ref{fig:extincurveerror} shows the average extinction curve from 49 curves weighted by the uncertainty, the shaded area represents 1$\sigma$ uncertainty in the weighted average, where the sigma is the median sigma of the corresponding 49 sources at the same wavelength. It can be seen that the effect of error on the extinction curve is apparent. The error also affects the silicate extinction profile parameters, which will be discussed in section 4.2. Considering that the extinction of silicate is closely related to the total extinction of stars, these sources are divided into two groups according to whether the $E(J-K_S)$ value is greater than 1.5. The average extinction curve weighted by the error is calculated  respectively, and shown on the right panel of Figure \ref{fig:extincurveerror}. The group with $E(J-K_S) < 1.5$ has smaller extinction with bigger error than that with $E(J-K_S) > 1.5$. When weighted by the uncertainty, sources with $E(J-K_S)>1.5$ have a larger average extinction in most infrared bands, but are slightly lower around the peak of 9.7$\mum$. Because the sources with smaller  $E(J-K_S)$ have larger errors,  their weights and influence are reduced in the average extinction curves so that the average curve closely resembles the case for $E(J-K_S)>1.5$.

\begin{figure*}
	\vspace{-1mm}
	\centering
	\includegraphics[width={18cm}]{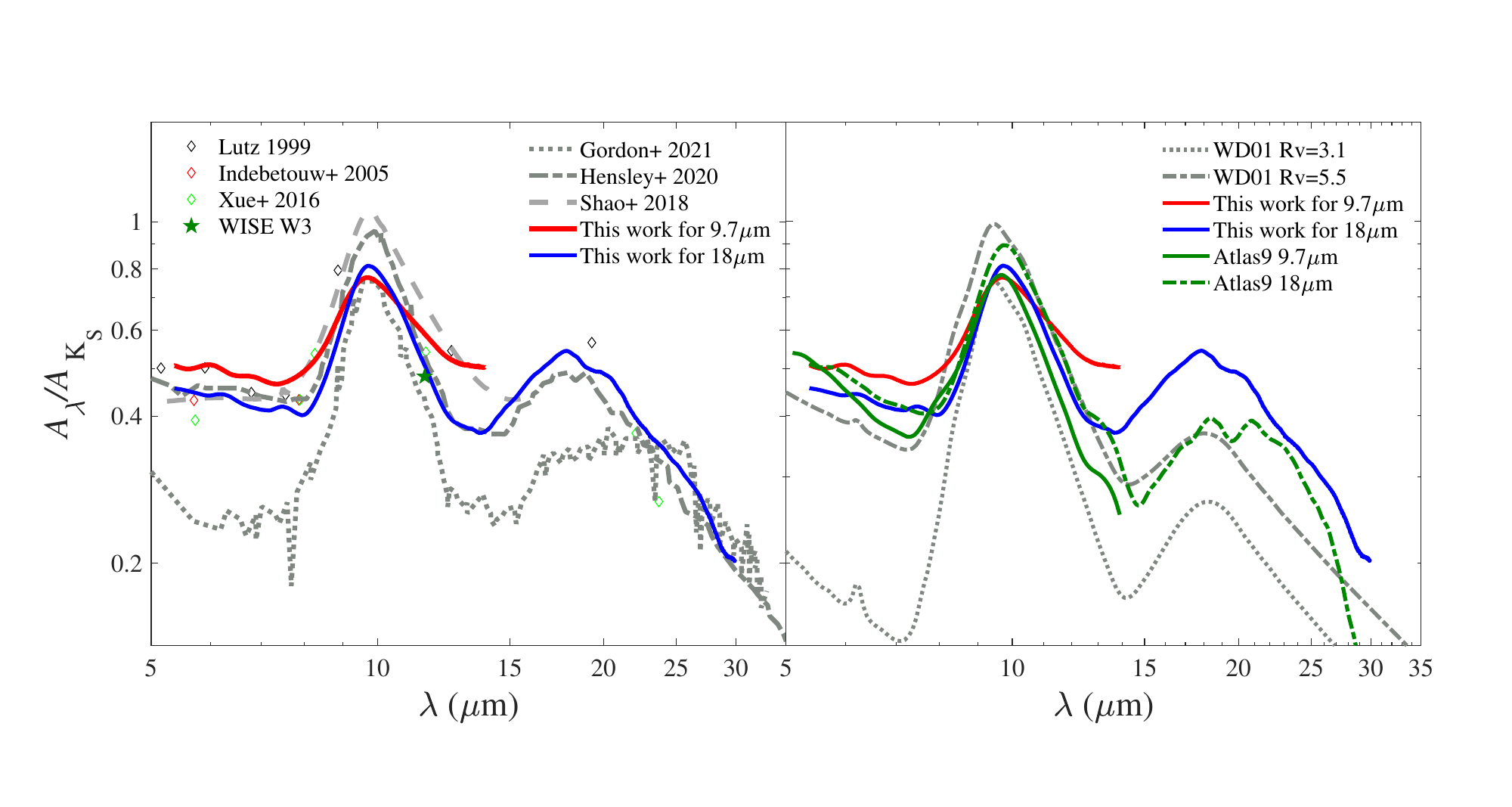}
	
	\caption{
		\label{fig:extincurvemean}
		Left: The mean extinction curve compared with \citet{hensley2020} and \citet{gordon2021}. Right: The mean extinction curve compared with that derived by using the stellar atmosphere model and interstellar dust model.
	}
	\vspace{4mm}
\end{figure*}

Figure \ref{fig:extincurvemean} shows the average extinction curves from the 49 and 6 curves respectively, weighted by $1/\sigma^2$ (The red and blue lines in Figure \ref{fig:extincurvemean}). The global trend agrees with the previously established curve. From 2$\mum$ on, the extinction decreases sharply with the wavelength until the local minimum at $\sim7\mum$, then it begins to increase because of the silicate feature and reaches the peak at  $\sim9.7\mum$, and the peak extinction of the silicate profile is comparable to that in the $K_S$ band. In general, the peak of the extinction curve is below our former result \citep{shao2018} and \citet{hensley2020}, but is consistent with \citet{gordon2021}, which also used the Spitzer/IRS spectrum with pair method. While \citet{gordon2021} choose 16 O-type or B-type stars,  our sample include all spectral-types. Early-type stars generally have a spectrum with fewer lines than late-type stars, but their continuum includes the free-free emission from ionized wind which needs to be treated very carefully \citep{hensley2020}. Between 5-7.5$\mum$, the extinction curve is higher than \citet{gordon2021}'s and consistent with the extinction value calculated from photometric data \citep{lutz1999,indebetouw2005,xue2016} and \citet{hensley2020}. On the other hand, the curve is below \citet{hensley2020}'s, especially around the peak region. It may be understood since the \citet{hensley2020} curve is derived from the observation of Cyg OB12-2, a highly buried star in Cygnus while our extinction law is the average of the 49 extinction curves. This difference may indicate the variation of the silicate extinction profile with sightline.


\subsubsection{The Profile around $18\mum$}

The extinction at the peak of the $18\mum$ feature is about half of the  $9.7\mum$ feature. This significant decrease in the intensity strength as well as the instrument sensitivity  leads to that only six extinction curves exhibit prominent  18 $\mum$ feature displayed in Figure \ref{fig:extincurve18}. As can be seen from this figure, the extinction profile of silicate dust around 18$\mum$ varies apparently. The peak wavelength fluctuates from 18 $\mum$ to 22 $\mum$. The extinction profile is more complicated, where the wavy shape is possibly caused by the low S/N. The average extinction curve from 6 curves in Figure \ref{fig:extincurvemean} weighted by $1/\sigma^2$ shows that the peak wavelength and the profile are similar to that of \citet{hensley2020}, but the peak value is a little bit lower.

\subsubsection{Comparison with the models}

As some works take the stellar model spectrum as the reference, we compared the results by using the observational spectrum of non-reddening star as the reference with using the theoretical spectrum by the ATLAS9 atmosphere model \citep{ATLAS9}. The comparison is displayed in the right panel of Figure \ref{fig:extincurvemean}. The extinction curves derived with stellar atmosphere model and with the observational spectrum are highly consistent at the 9.7$\mum$ and lower around  18$\mum$.  A visible difference is that the curve from stellar model  shows a more prominent feature at $\sim13.2\mum$. The agreement indicates that there is no significant difference in taking the stellar model or the observed spectrum as the reference.

The extinction curves are also compared  with that derived from the dust model by \citet{WD01}. As shown in the right panel of Figure \ref{fig:extincurvemean},  the curve is basically consistent with the WD01 curve with $R_V$=5.5 at 9.7$\mum$ and higher around 18$\mum$, but much higher than that with $R_V$=3.1. This agrees with the conclusions of previous studies of the average infrared extinction law (e.g. \citealt{xue2016}).

Regardless of whether the pair method or an atmospheric model method, we can find a tiny bump feature in the range of 12.7-14.1 $\mum$. \citet{vandishoeck2023} reported two broad bumps centered at 7.7 and 13.7$\mum$ from proto-planetary disk by JWST, which are due to the $\nu 4 + \nu5$ and $\nu5$ bands of C$_2$H$_2$ \citep{tabone2023}. It seems that this tiny bump coincides with the C$_2$H$_2$ feature around 13.7$\mum$, but the lack of the 7.7$\mum$ feature challenges such identification. The JWST/MIRI spectra in McClure et al (2023)  show a broad absorption feature near 11.5$\mum$, which can plausibly be attributed to libration in H$_2$O ice. The extension of this feature to long wavelength may partially explain the tiny bump feature at 12.7-14.1$\mum$, but the spectra have low S/N and this feature can hardly extend to 14$\mum$. The nature of this feature needs further study.





\subsection{Relation of the Peak and the FWHM of the 9.7$\mum$ Feature}

\begin{figure*}
	\vspace{-1mm}
	\centering
	\includegraphics[width={12cm}]{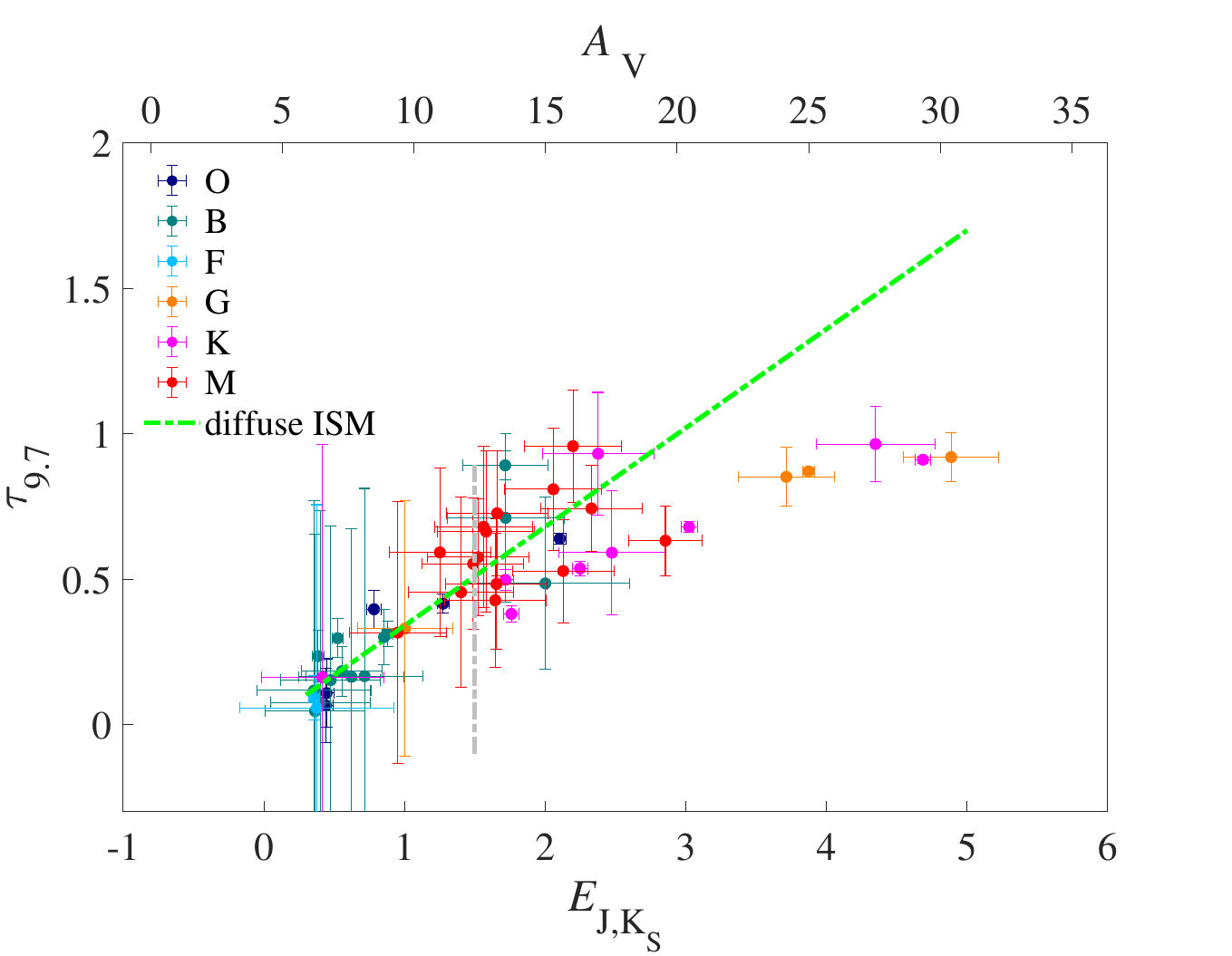}
	
	\caption{
		\label{fig:ejktau97}
		The relation between the color excess $E(J-K_S)$, the optical extinction at V band $A_{\rm V}$ and the optical depth at 9.7$\mum$ $\Delta\tau_{9.7}$. The color denotes the spectral type of stars. The green dash line is the relation in diffuse ISM. The gray dash line is the reference line at $E(J-K_S)$=1.5.
	}
	\vspace{4mm}
\end{figure*}

The dependence of the silicate profile on the color excess is examined since the silicate grains may be different with the environment density or temperature or radiation field. Figure \ref{fig:ejktau97} presents the silicate optical depth at 9.7$\mum$ and the color excess $E_{\rm JK_S}$. The silicate optical depth increases with $E_{\rm JK_S}$ as expected since both are proportional to the total extinction. The whole trend follows the relation of the diffuse medium,  which coincides with the result of \citet{chiar2007} for diffuse medium (The green dashed line in Figure \ref{fig:ejktau97}). However, when $E_{\rm JK_S}$ becomes larger than 4, the silicate optical depth at 9.7$\mum$ does not seem to increase with the color excess. This may imply that the stars with large color excess have circumstellar silicate emission to reduce the 9.7$\mum$ extinction in spite of that we already tried to remove the stars with excess emission in the WISE/W3 band. \citet{chiar2007} found similar bifurcating, and they attributed it to the effect of dust grain growth in the high extinction regions. The dust grain growth increases the near-IR extinction while has no effect on the silicate feature.

The optical extinction in the V band was found to be proportional to the optical depth at 9.7$\mum$ for the diffuse interstellar medium, quantitatively  $A_{\rm V}/\Delta\tau_{9.7}$ $\approx$ 18.  The value of $A_{\rm V}$ of the targets is calculated by the relation $E_{\rm JK_S}/A_{\rm V}$ $\approx$ 0.165 \citep{chiar2007} and labeled by the upper abscissa in Figure \ref{fig:ejktau97}. Consequently, this line implies the ratio $A_{\rm V}/\Delta\tau_{9.7}$ $\approx$ 17.83, consistent with previous results. It should be mentioned that this ratio depends on the conversion relation between $E_{\rm JK_S}$ and $A_{\rm V}$, and can differ with the conversion factor. On the other hand, this ratio increases when the extinction at V band $A_{\rm V}$ is particularly large. One possible reason is mentioned above, i.e. this group is consisted of late-type stars with circumstellar silicate emission that fills up the absorption. Another possible reason is the dust size effect as suggested by \citet{chiar2007}. The $A_{\rm V}/\Delta\tau_{9.7}$ ratio for pure silicate dust is $\sim$ 1 for very small (a$\sim$ 0.01$\mum$) particles, rising to $\sim$ 5 for classical (a$\sim$ 0.15$\mum$) grains (e.g. \citealp{stephens1980,gillett1975,shao2018}). By adding carbonaceous dust with $M_{\rm carb}/M_{\rm sil} \textgreater 0.15$, this ratio can rise to 18 \citep{shao2017}. So another possible reason for the larger ratio $A_{\rm V}/\Delta\tau_{9.7}$  is that the medium in these sightlines contains more carbon dust.

\begin{figure*}
	\vspace{-1mm}
	\centering
	\includegraphics[width={12.0cm}]{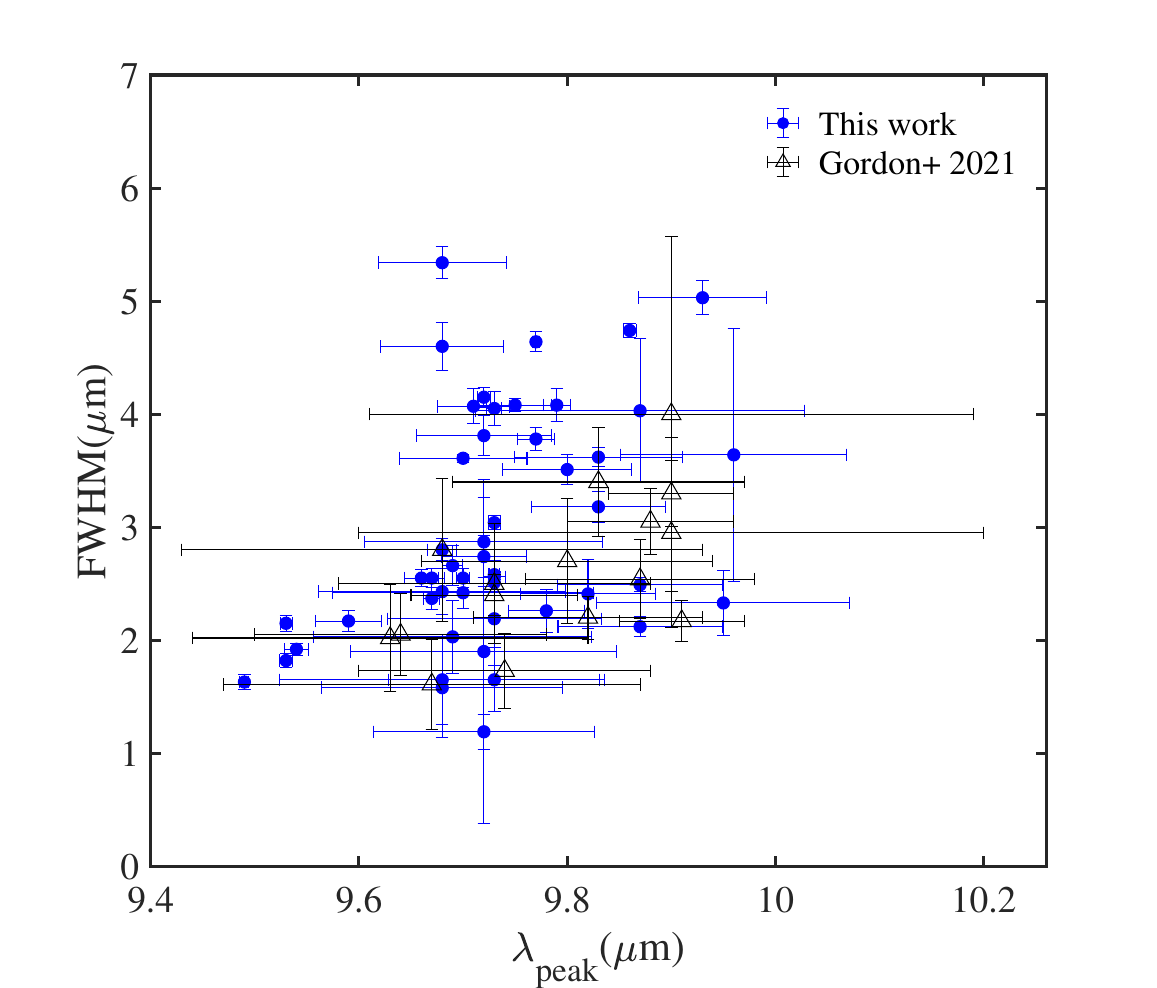}
	
	
	\caption{
		\label{fig:gordon21}
		The FWHM and peak wavelength of the 9.7$\mum$ silicate profile from this work compared with \citet{gordon2021}.
	}
	\vspace{4mm}
\end{figure*}

On the correlation between the peak wavelength and the FWHM of the feature,  \citet{gordon2021} thought they are well correlated. In Figure \ref{fig:gordon21}, our data are compared with theirs, which present similar scattering.
The correlation coefficient is 0.4, which indicates a positive correlation considering the uncertainties of the parameters. It should be noted that our peaks and FWHMs are measured directly from the curve after subtracting the continuum extinction (mentioned at the end of Section 2), and no Drude or Gaussian profile fitting is performed. This study has 6 stars in common with \citet{gordon2021}, namely VI CYG 1, HD229238, VI CYG 2, HD147889, HD147701 and HD283809. In  comparing the profile parameters, the peak wavelengths are essentially similar after accounting for the errors, while the difference in FWHM is obvious. With the exception of VI CYG 1 for which the FWHM is consistent, there is a difference of 1-2 $\mum$ in the FWHM of the other five sources. This is partly due to the fact that the peak and FWHM themselves have some errors, whether measured by us or \citet{gordon2021}, and partly probably due to the spectral data limitations, including the choice of different reference stars, and the differences in the method of subtracting continuum extinction and measuring the profile widths.
It is worth noting that the FWHM for \citet{gordon2021} in our own measurement is based on the extinction curve provided by \citet{gordon2021}, not the $\gamma_0$ given in Table 6 of  \citet{gordon2021}. We found the difference between the FWHM and  $\gamma_0$ to be very small.

\begin{figure*}
	\vspace{-1mm}
	\centering
	\includegraphics[width={16.0cm}]{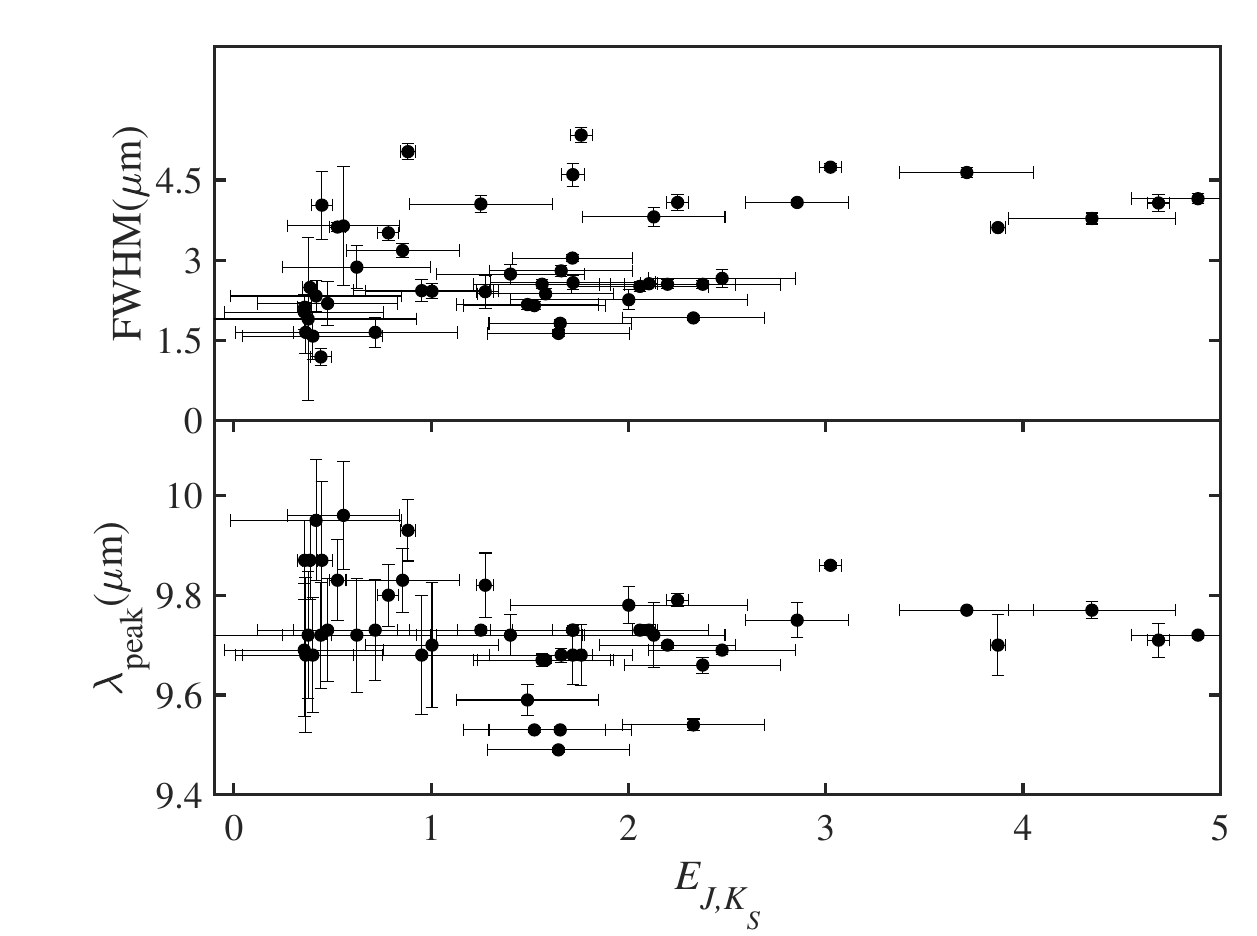}
	
	
	\caption{
		\label{fig:peakejk}
		The relationship between the $E(J-K_S)$ with FWHM and peak wavelength of the 9.7$\mum$ silicate profile.
	}
	\vspace{4mm}
\end{figure*}

Considering correlation between $E_{\rm JK_S}$ and $\Delta\tau_{9.7}$, the effect of $E_{\rm JK_S}$ on the peak and FWHM of the extinction profile at 9.7$\mum$ is further investigated and is shown in Figure \ref{fig:peakejk}, and no significant correlation is present.

\subsection{Dependence of the Silicate Profile Features on Stellar Spectral Type}

\begin{figure*}
	\vspace{-1mm}
	\centering
	\includegraphics[width={17.0cm}]{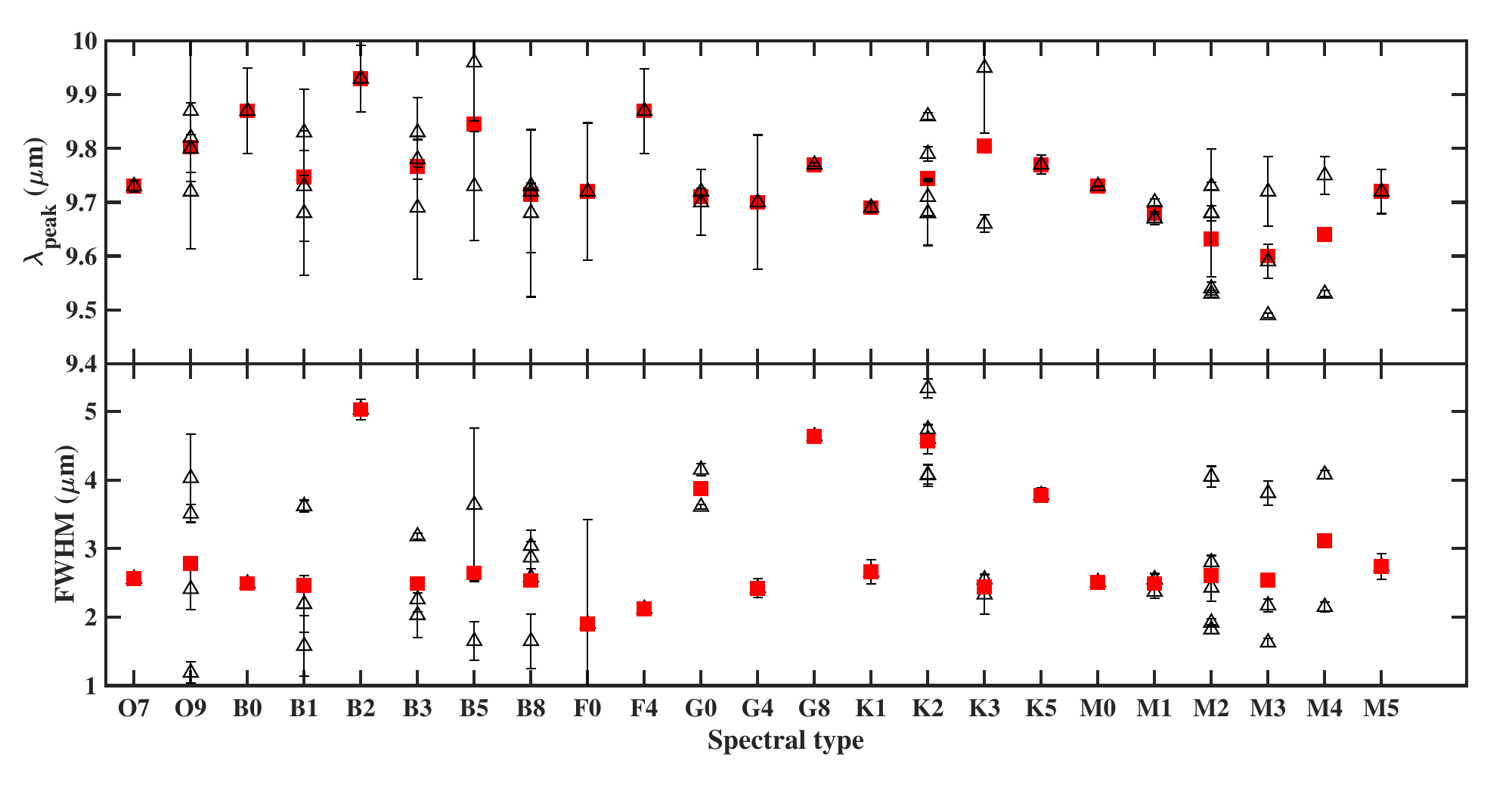}
	
	
	\caption{
		\label{fig:spdis1}
	The variation of the FWHM and peak wavelength of the 9.7$\mum$ silicate profile with the spectral type of the tracing stars.
	The black triangle with the error bar represents the peak value or FWHM of each source, and the red square represents the mean of the corresponding source at the same spectral type.}
	\vspace{4mm}
\end{figure*}

Figure \ref{fig:spdis1} shows that the peak wavelength of the 9.7$\mum$ profile may be divided into two groups. For OB- to even early F-type stars, the peak concentrates at longer wavelength around 9.7-9.8$\mum$ with the maximum being 9.9$\mum$, while it moves to about 9.6$\mum$ for late F- to GKM-type stars with the minimum being 9.5$\mum$. It may be caused by that the crystalline silicate around late-type stars shifts the peak wavelength since the crystalline silicate emits stronger at $\lambda> 9.7\mum$ \citep{chen2016}. \citet{do2020} found a minor absorption band around 11.1 $\mum$ whose carrier is attributed to crystalline forsterite. This weak feature is also found in a small fraction of our sources, such as HD 283809, VI CYG 1, HD 29647, CI*IC 348 LRL 11. This drop is not so obvious after considering the error, and whether this trend is real needs further study. On the other hand, the FWHM shows no systematic dependence on spectral type as exhibited in the same figure that the variation of FWHM with spectral type is irregular. The most narrow FWHM is about 1.2$\mum$, occurring in an O9-type star, and the widest is about 5$\mum$ seen in a few stars, meanwhile, most FWHM are between 2.0 and 3$\mum$.

Detailed experimental studies have revealed the existence of tight correlations of silicate profile properties with the dust physical and chemical parameters. \citet{koike1987} investigated the infrared spectra of grains of natural and synthesized silicate glasses covering a wide range of SiO$_2$. They found that the peak wavelength and the width of the 10 $\mum$ band decrease with increasing SiO$_2$, whereas the peak absorption of the band increases with growing SiO$_2$. Dust extinction originates from molecular vibration, especially the chemical bond stretching or bending. The influence factors of the dust extinction profile include dust composition, size distribution, and shape. Based on our research as well as others （e.g. \citealt{shao2018}), the variety of dust composition may account for the above mentioned variation of the profile.


\section{Summary}

The Spitzer/IRS spectra as well as the 2MASS and WISE photometry of 49 stars with obvious extinction are analyzed to obtain the extinction curves in the infrared by using the "pair" method. The silicate extinction profile around 9.7$\mum$ and 18$\mum$ are determined after subtracting the continuum by a constant power-law function. The characteristic parameters of the 9.7$\mum$ profile are calculated and investigated. The following results are obtained:

(i) The general 9.7$\mum$ and 18$\mum$ silicate extinction features peak around $\simali$9.2- 9.8$\mum$ and $\simali$ 18-22$\mum$ respectively.

(ii) The wavelength of the peak of the silicate extinction profile may decreases with the spectral type, but this trend is not so clear due to the presence of errors and needs to be verified with more data support in the future, meanwhile the FWHM is independent of the spectral type.

(iii) For most sources, the silicate optical depth increases with $E_{\rm JK_S}$, consistent with the proportional relation found in diffuse medium. However, when $E_{\rm JK_S} \textgreater $4, this relationship is not applicable. Some late-type stars may contain circumstellar emission, weakening the silicate optical depth at 9.7$\mum$. The mean ratio of the visual extinction to the 9.7$\mum$ silicate absorption optical depth for diffuse medium is 17.83, in close agreement with that of the solar neighborhood diffuse ISM. Still, for some late-type stars, the ratio is larger than that of the neighborhood diffuse ISM, which may be caused by the silicate dust emission from the circumstellar dust.


\section{Acknowledgements}

We are grateful to Drs Yi Ren, Jun Li, Mao Yuan and Shu Wang for their helpful discussion. We also thank the anonymous referee for his/her helpful suggestions. This work is supported by the Beijing Academy of Science and Technology project BGS202205, BGS202105, the NSFC project 12133002, National Key R\&D Program of China No. 2019YFA0405503, and CMS-CSST-2021-A09. This work has made use of the data from Spitzer, 2MASS, WISE and Gaia.


\bibliography{szzref}{}

\begin{thebibliography}{}
\expandafter\ifx\csname natexlab\endcsname\relax\def\natexlab#1{#1}\fi
\providecommand{\url}[1]{\href{#1}{#1}}
\providecommand{\dodoi}[1]{doi:~\href{http://doi.org/#1}{\nolinkurl{#1}}}
\providecommand{\doeprint}[1]{\href{http://ascl.net/#1}{\nolinkurl{http://ascl.net/#1}}}
\providecommand{\doarXiv}[1]{\href{https://arxiv.org/abs/#1}{\nolinkurl{https://arxiv.org/abs/#1}}}

\bibitem[{Allen \& Cox(2000)}]{allen2000}
Allen, C.~W., \& Cox, A.~N. 2000, Allen's astrophysical quantities (Springer
  Science \& Business Media)

\bibitem[{Castelli {et~al.}(1997)Castelli, Gratton, \& Kurucz}]{castelli1997}
Castelli, F., Gratton, R., \& Kurucz, R. 1997, Astronomy and Astrophysics, 318,
  841

\bibitem[{Chen {et~al.}(2016)Chen, Luo, Liu, \& Jiang}]{chen2016}
Chen, R., Luo, A., Liu, J., \& Jiang, B. 2016, The Astronomical Journal, 151,
  146

\bibitem[{Chiar \& Tielens(2006)}]{chiar2006}
Chiar, J., \& Tielens, A. 2006, The Astrophysical Journal, 637, 774

\bibitem[{Chiar {et~al.}(2007)Chiar, Ennico, Pendleton, Boogert, Greene, Knez,
  Lada, Roellig, Tielens, Werner, {et~al.}}]{chiar2007}
Chiar, J., Ennico, K., Pendleton, Y., {et~al.} 2007, The Astrophysical Journal,
  666, L73

\bibitem[{Do-Duy {et~al.}(2020)Do-Duy, Wright, Fujiyoshi, Glasse, Siebenmorgen,
  Smith, Stecklum, \& Sterzik}]{do2020}
Do-Duy, T., Wright, C.~M., Fujiyoshi, T., {et~al.} 2020, Monthly Notices of the
  Royal Astronomical Society, 493, 4463

\bibitem[{Fogerty {et~al.}(2016)Fogerty, Forrest, Watson, Sargent, \&
  Koch}]{fogerty2016}
Fogerty, S., Forrest, W., Watson, D.~M., Sargent, B.~A., \& Koch, I. 2016, The
  Astrophysical Journal, 830, 71

\bibitem[{Gillett {et~al.}(1975)Gillett, Merrill, \& Soifer}]{gillett1975}
Gillett, F., Merrill, K., \& Soifer, B. 1975, Astrophysical Journal, vol. 200,
  Sept. 15, 1975, pt. 1, p. 609-620., 200, 609

\bibitem[{Gordon {et~al.}(2023)Gordon, Clayton, Decleir, Fitzpatrick, Massa,
  Misselt, \& Tollerud}]{gordon2023one}
Gordon, K.~D., Clayton, G.~C., Decleir, M., {et~al.} 2023, The Astrophysical
  Journal, 950, 86

\bibitem[{Gordon {et~al.}(2021)Gordon, Misselt, Bouwman, Clayton, Decleir,
  Hines, Pendleton, Rieke, Smith, \& Whittet}]{gordon2021}
Gordon, K.~D., Misselt, K.~A., Bouwman, J., {et~al.} 2021, The Astrophysical
  Journal, 916, 33

\bibitem[{Henning(2010)}]{henning2010}
Henning, T. 2010, Annual Review of Astronomy and Astrophysics, 48, 21

\bibitem[{Hensley \& Draine(2020)}]{hensley2020}
Hensley, B.~S., \& Draine, B. 2020, The Astrophysical Journal, 895, 38

\bibitem[{Houck {et~al.}(2004)Houck, Roellig, Van~Cleve, Forrest, Herter,
  Lawrence, Matthews, Reitsema, Soifer, Watson, {et~al.}}]{houck2004}
Houck, J.~R., Roellig, T.~L., Van~Cleve, J., {et~al.} 2004, The Astrophysical
  Journal Supplement Series, 154, 18

\bibitem[{Indebetouw {et~al.}(2005)Indebetouw, Mathis, Babler, Meade, Watson,
  Whitney, Wolff, Wolfire, Cohen, Bania, {et~al.}}]{indebetouw2005}
Indebetouw, R., Mathis, J., Babler, B., {et~al.} 2005, The Astrophysical
  Journal, 619, 931

\bibitem[{Kemper {et~al.}(2004)Kemper, Vriend, \& Tielens}]{kemper2004}
Kemper, F., Vriend, W., \& Tielens, A. 2004, The Astrophysical Journal, 609,
  826

\bibitem[{Koike \& Hasegawa(1987)}]{koike1987}
Koike, C., \& Hasegawa, H. 1987, Astrophysics and space science, 134, 361

\bibitem[{Kurucz(2014)}]{ATLAS9}
Kurucz, R.~L. 2014, Determination of Atmospheric Parameters of B-, A-, F-and
  G-Type Stars: Lectures from the School of Spectroscopic Data Analyses, 39

\bibitem[{Li \& Draine(2001)}]{li2001}
Li, A., \& Draine, B. 2001, The Astrophysical Journal, 550, L213

\bibitem[{Liu {et~al.}(2017)Liu, Jiang, Li, \& Gao}]{liu2017}
Liu, J., Jiang, B., Li, A., \& Gao, J. 2017, Monthly Notices of the Royal
  Astronomical Society, 466, 1963

\bibitem[{Lutz(1999)}]{lutz1999}
Lutz, D. 1999, in The Universe as Seen by ISO, Vol. 427, 623

\bibitem[{McClure(2009)}]{mcclure2009}
McClure, M. 2009, The Astrophysical Journal, 693, L81

\bibitem[{Min {et~al.}(2007)Min, Waters, de~Koter, Hovenier, Keller, \&
  Markwick-Kemper}]{min2007}
Min, M., Waters, L., de~Koter, A., {et~al.} 2007, Astronomy \& Astrophysics,
  462, 667

\bibitem[{Molster \& Kemper(2005)}]{molster2005}
Molster, F., \& Kemper, C. 2005, Space Science Reviews, 119, 3

\bibitem[{Olofsson \& Olofsson(2011)}]{olofsson2011}
Olofsson, S., \& Olofsson, G. 2011, Astronomy \& Astrophysics, 534, A127

\bibitem[{Shao {et~al.}(2017)Shao, Jiang, \& Li}]{shao2017}
Shao, Z., Jiang, B., \& Li, A. 2017, The Astrophysical Journal, 840, 27

\bibitem[{Shao {et~al.}(2018)Shao, Jiang, Li, Gao, Lv, \& Yao}]{shao2018}
Shao, Z., Jiang, B., Li, A., {et~al.} 2018, Monthly Notices of the Royal
  Astronomical Society, 478, 3467

\bibitem[{Stephens(1980)}]{stephens1980}
Stephens, J.~R. 1980, Astrophysical Journal, Part 1, vol. 237, Apr. 15, 1980,
  p. 450-461., 237, 450

\bibitem[{Tabone {et~al.}(2023)Tabone, Bettoni, van Dishoeck, Arabhavi, Grant,
  Gasman, Henning, Kamp, G{\"u}del, Lagage, {et~al.}}]{tabone2023}
Tabone, B., Bettoni, G., van Dishoeck, E., {et~al.} 2023, Nature Astronomy, 1

\bibitem[{Van~Breemen {et~al.}(2011)Van~Breemen, Min, Chiar, Waters, Kemper,
  Boogert, Cami, Decin, Knez, Sloan, {et~al.}}]{vanbreemen2011}
Van~Breemen, J., Min, M., Chiar, J., {et~al.} 2011, Astronomy \& Astrophysics,
  526, A152

\bibitem[{van Dishoeck {et~al.}(2023)van Dishoeck, Grant, Tabone, van Gelder,
  Francis, Tychoniec, Bettoni, Arabhavi, Gasman, Kavanagh,
  {et~al.}}]{vandishoeck2023}
van Dishoeck, E., Grant, S., Tabone, B., {et~al.} 2023, Faraday Discussions

\bibitem[{Weingartner \& Draine(2001)}]{WD01}
Weingartner, J.~C., \& Draine, B.~T. 2001, The Astrophysical Journal, 548, 296,
  \dodoi{10.1086/318651}

\bibitem[{Werner {et~al.}(2004)Werner, Roellig, Low, Rieke, Rieke, Hoffmann,
  Young, Houck, Brandl, Fazio, {et~al.}}]{werner2004spitzer}
Werner, M.~W., Roellig, T., Low, F., {et~al.} 2004, The Astrophysical Journal
  Supplement Series, 154, 1

\bibitem[{Xue {et~al.}(2016)Xue, Jiang, Gao, Liu, Wang, \& Li}]{xue2016}
Xue, M., Jiang, B., Gao, J., {et~al.} 2016, The Astrophysical Journal
  Supplement Series, 224, 23

\end{thebibliography}
\bibliographystyle{aasjournal}



\clearpage

%

\clearpage


	\begin{footnotesize}
		\begin{longtable}{l|l|c|c|c|c|c|c|c}
			\caption{Stellar parameters for the target and reference stars}\label{table:t1} \\
			
			\hline
			\hline
			Target Object &Type & RA & DEC  &  Reference Object &Type & RA & DEC& $E(J-K_S)$\\
			& &(degrees) & (degrees) &  &  & (degrees) &(degrees)&(mag) \\
			\hline
			\endfirsthead
			\caption{Stellar parameters for the target and reference stars}\\
			\hline
			\hline
			Target Object &Type & RA & DEC  &  Reference Object &Type & RA & DEC& $E(J-K_S)$\\
			& &(degrees) & (degrees) &  &  & (degrees) &(degrees)&(mag) \\
			\hline
			\endhead
			\hline \endfoot
			StRS 354	&	O7-B3I	&	307.31 	&	37.85 	&	HD 188209	&	O9.5Iab	&	298.00 	&	47.03 	&	2.10 \\
			HD 152424	&	O9.2I	&	253.76 	&	-42.09	&	HD 188209	&	O9.5Iab	&	298.00 	&	47.03 	&	0.44 \\
			V* V1074 Sco	&	O9.2I	&	256.62 	&	-35.45	&	HD 188209	&	O9.5Iab	&	298.00 	&	47.03 	&	0.45 \\
			IC 1795 109	&	O9.7I	&	36.69 	&	62.05 	&	HD 188209	&	O9.5Iab	&	298.00 	&	47.03 	&	1.27 \\
			VI CYG 1	&	O9V	&	307.79 	&	41.53 	&	* 10 Lac	&	O9V	&	339.82 	&	39.05 	&	0.78 \\
			HD 229238	&	B0.5I	&	306.02 	&	38.54 	&	HR 3091	&	B2V	&	118.55 	&	-35.88	&	0.39 \\
			HD 14956	&	B1.5I	&	36.69 	&	57.68 	&	* tet Ara	&	B2Ib	&	271.66 	&	-50.09	&	0.40 \\
			V* DW Cru	&	B1I	&	194.14 	&	-64.36	&	* tet Ara	&	B2Ib	&	271.66 	&	-50.09	&	0.47 \\
			VI CYG 2	&	B1Ib	&	307.84 	&	41.52 	&	HR 3091	&	B2V	&	118.55 	&	-35.88	&	0.52 \\
			HD 147889	&	B2III	&	246.35 	&	-24.47	&	HR 3091	&	B2V	&	118.55 	&	-35.88	&	0.88 \\
			VI CYG 12	&	B3-4I	&	308.17 	&	41.24 	&	* tet Ara	&	B2Ib	&	271.66 	&	-50.09	&	2.00 \\
			HD 157038	&	B3I	&	260.66 	&	-37.80	&	* tet Ara	&	B2Ib	&	271.66 	&	-50.09	&	0.36 \\
			BD+43 3710	&	B5I	&	311.39 	&	43.54 	&	* h Car	&	B5II	&	143.61 	&	-59.23	&	0.72 \\
			HD 147701	&	B5III	&	246.09 	&	-25.03	&	* iot Aql	&	B5III	&	294.18 	&	-1.29	&	0.56 \\
			StRS 136	&	B8-A9I	&	266.98 	&	-29.20	&	* h Car	&	B5II	&	143.61 	&	-59.23	&	1.72 \\
			StRS 164	&	B8-A9I	&	274.08 	&	-16.60	&	* h Car	&	B5II	&	143.61 	&	-59.23	&	1.72 \\
			V* V965 Cen	&	B8I	&	200.73 	&	-62.01	&	* tet Ara	&	B2Ib	&	271.66 	&	-50.09	&	0.36 \\
			HD 29647	&	B8III	&	70.28 	&	25.99 	&	* eps Sgr	&	B9.5III	&	276.04 	&	-34.38	&	0.62 \\
			* phi Cas	&	F0I	&	20.02 	&	58.23 	&	* 22 And	&	F5II	&	2.58 	&	46.07 	&	0.38 \\
			HD 227575	&	F4I	&	301.39 	&	33.74 	&	HR 8441	&	F1IV	&	332.07 	&	25.54 	&	0.36 \\
			HD 283809	&	F8III	&	70.35 	&	25.91 	&	V* S TrA	&	F8II	&	240.29 	&	-63.78	&	0.40 \\
			
			TIC 415790729&	G04III	&	277.44 	&	1.31 	&	HD 39833	&	G0III	&	88.76 	&	-0.51	&	3.87 \\
			TIC 268324610&	G0-M3III	&	68.06 	&	24.49 	&	HD 161178	&	G9III	&	264.29 	&	72.46 	&	4.80 \\
			Cl* IC 348 LRL 11	&	G4III	&	56.28 	&	32.07 	&	HD 161178	&	G9III	&	264.29 	&	72.46 	&	0.85\\
			TIC 345429046&	G8.5III	&	326.55 	&	47.58 	&	HD 161178	&	G9III	&	264.29 	&	72.46 	&	3.72 \\
			JHR 80 D-14	&	K1III	&	187.91 	&	-63.79	&	* 51 Her	&	K2Iab:	&	252.94 	&	24.66 	&	2.36 \\
			TIC 345611668	&	K2III	&	326.84 	&	47.63 	&	HD 52418	&	K2III	&	103.92 	&	-68.51	&	1.72 \\
			Elia 3-3	&	K2III	&	65.85 	&	25.00 	&	HD 52418	&	K2III	&	103.92 	&	-68.51	&	1.76 \\
			Elia 3-13	&	K2III	&	68.36 	&	26.26 	&	HD 52418	&	K2III	&	103.92 	&	-68.51	&	2.25 \\
			TIC 190930040	&	K2III	&	326.14 	&	47.58 	&	HD 52418	&	K2III	&	103.92 	&	-68.51	&	3.02 \\
			Elia 3-16	&	K2III	&	69.91 	&	26.19 	&	HD 52418	&	K2III	&	103.92 	&	-68.51	&	4.69 \\
			2MASS J18391505-0605191	&	K3I	&	279.81 	&	-6.09	&	* 51 Her	&	K2Iab:	&	252.94 	&	24.66 	&	2.38 \\
			GBM90 Per 75	&	K3III	&	56.28 	&	32.57 	&	HD 42701	&	K3III	&	91.71 	&	-67.28	&	0.42 \\
			TAMURA 8	&	K5III	&	70.24 	&	25.90 	&	HD 53561	&	K5III	&	106.52 	&	13.99 	&	4.11 \\
			TIC 328138536	&	M0I	&	279.84 	&	-6.04	&	HD 13570	&	M0III	&	32.56 	&	-61.10	&	1.56 \\
			IRAS 20315+4026	&	M0Iab:	&	308.35 	&	40.61 	&	HD 13570	&	M0III	&	32.56 	&	-61.10	&	2.06 \\
			TIC 328281708	&	M1I	&	279.85 	&	-5.98	&	HD 13570	&	M0III	&	32.56 	&	-61.10	&	1.58 \\
			TIC 7759115	&	M1I	&	279.79 	&	-6.02	&	HD 13570	&	M0III	&	32.56 	&	61.10	&	2.20 \\
			StRS 234	&	M2I	&	279.84 	&	-6.03	&	TYC 3440-13-1	&	M3.0V	&	151.09 	&	50.39 	&	1.65 \\
			Cl Stephenson 2 9	&	M2I	&	279.85 	&	-6.01	&	HD 13570	&	M0III	&	32.56 	&	-61.10	&	1.66 \\
			TIC 328281041	&	M2I	&	279.82 	&	-6.03	&	HD 13570	&	M0III	&	32.56 	&	-61.10	&	2.33 \\
			Cl* NGC 7419 BMD 921	&	M2Iab	&	343.63 	&	60.80 	&	HD 13570	&	M0III	&	32.56 	&	-61.10	&	0.95 \\
			WOS 35	&	M2III	&	192.14 	&	-62.08	&	HD 189246	&	M2III	&	300.17 	&	-40.20	&	1.03 \\
			Cl Stephenson 2 3	&	M3I	&	279.83 	&	-5.99	&	HD 13570	&	M0III	&	32.56 	&	-61.10	&	1.49 \\
			Cl Stephenson 2 8	&	M3I	&	279.85 	&	-6.05	&	TYC 3440-13-1	&	M3.0V	&	151.09 	&	50.39 	&	1.65 \\
			JHR 80 D-12	&	M3III	&	187.87 	&	-63.71	&	HD 189246	&	M2III	&	300.17 	&	-40.20	&	1.89 \\
			Cl Stephenson 2 7	&	M4I	&	279.84 	&	-6.00	&	HD 13570	&	M0III	&	32.56 	&	-61.10	&	1.52 \\
			TIC 150058734	&	M4III	&	69.86 	&	25.88 	&	V* VX Pic	&	M4III	&	75.75 	&	-54.10	&	2.86 \\
			JHR 80 D-11	&	M5III	&	187.85 	&	-63.68	&	V* V464 Car	&	M5III	&	122.20 	&	-61.57	&	1.40 \\

		\end{longtable}
	\end{footnotesize}

\clearpage


\begin{footnotesize}
	\begin{longtable}{l|c|c|c|c|c|c|c}
	\caption{Infrared photometry for the target and reference stars }\label{table:t2}\\
			
		\hline
		\hline
				Object & J & H & $\rm \Ks$&W1& W2 & W3 & W4 \\
				& 1.235$\mum$ &1.662$\mum$  &2.159$\mum$ & 3.353$\mum$ &4.603$\mum$  &11.561$\mum$  &22.088$\mum$\\
				& (mag)& (mag) & (mag) & (mag)& (mag) & (mag) & (mag)  \\
				\hline
				\endfirsthead
				\caption{IR photometry for the target and reference stars } \\
				\hline
				\hline
				Object & J & H & $\rm \Ks$&W1& W2 & W3 & W4 \\
				& 1.235$\mum$ &1.662$\mum$  &2.159$\mum$ & 3.353$\mum$ &4.603$\mum$  &11.561$\mum$  &22.088$\mum$\\
				& (mag)& (mag) & (mag) & (mag)& (mag) & (mag) & (mag)  \\
				\hline
				
				\endhead
				\hline \endfoot
				StRS 354	&	7.62 	&	6.37 	&	5.74 	&	5.28 	&	4.80 	&	5.08 	&	4.60 \\
				V* V1074 Sco	&	5.02 	&	4.85 	&	4.75 	&	4.43 	&	3.74 	&	4.57 	&	4.21 \\
				IC 1795 109	&	8.37 	&	7.68 	&	7.26 	&	7.04 	&	6.87 	&	7.25 	&	3.91 \\
				VI CYG 1	&	7.97 	&	7.56 	&	7.37 	&	7.21 	&	7.19 	&	7.31 	&	7.96 \\
				HD 152424	&	5.32 	&	5.22 	&	5.06 	&	4.97 	&	4.57 	&	4.86 	&	4.51 \\
				HD 229238	&	7.02 	&	6.82 	&	6.72 	&	6.64 	&	6.64 	&	6.59 	&	5.70 \\
				HD 14956	&	5.68 	&	5.47 	&	5.35 	&	5.34 	&	4.97 	&	5.14 	&	4.84 \\
				V* DW Cru	&	5.57 	&	5.36 	&	5.18 	&	5.07 	&	4.71 	&	4.94 	&	4.65 \\
				VI CYG 2	&	8.08 	&	7.75 	&	7.63 	&	7.47 	&	7.45 	&	7.54 	&	7.30 \\
				HD 147889	&	5.34 	&	4.94 	&	4.58 	&	4.39 	&	3.71 	&	4.35 	&	1.89 \\
				VI CYG 12	&	4.67 	&	3.51 	&	2.70 	&	2.67 	&	1.94 	&	2.23 	&	1.76 \\
				HD 157038	&	4.45 	&	4.37 	&	4.14 	&	3.90 	&	3.12 	&	3.81 	&	3.60 \\
				BD+43 3710	&	6.60 	&	6.14 	&	5.88 	&	5.70 	&	5.45 	&	5.48 	&	4.02 \\
				HD 147701	&	6.67 	&	6.38 	&	6.19 	&	6.01 	&	6.00 	&	6.14 	&	5.89 \\
				StRS 136	&	6.88 	&	5.73 	&	5.10 	&	5.43 	&	4.67 	&	4.63 	&	3.92 \\
				StRS 164	&	7.66 	&	6.50 	&	5.89 	&	5.67 	&	4.99 	&	5.52 	&	6.42 \\
				V* V965 Cen	&	6.12 	&	5.88 	&	5.70 	&	5.64 	&	5.32 	&	5.47 	&	5.19 \\
				HD 29647	&	5.96 	&	5.59 	&	5.36 	&	5.20 	&	4.86 	&	4.93 	&	3.82 \\
				* phi Cas	&	3.75 	&	3.54 	&	3.19 	&	2.92 	&	2.13 	&	2.82 	&	2.69 \\
				HD 227575	&	6.37 	&	6.01 	&	5.74 	&	5.55 	&	5.32 	&	5.52 	&	5.28 \\
				HD 283809	&	6.99 	&	6.46 	&	6.17 	&	5.99 	&	5.84 	&	6.02 	&	5.78 \\
				TIC 415790729	&	12.87 	&	9.94 	&	8.54 	&	7.93 	&	7.55 	&	7.55 	&	7.48 \\
				TIC 268324610	&	13.36 	&	9.91 	&	8.12 	&	7.45 	&	6.76 	&	6.77 	&	6.40 \\
				Cl* IC 348 LRL 11	&	10.28 	&	9.32 	&	8.88 	&	8.60 	&	8.49 	&	8.30 	&	7.27 \\
				TIC 345429046	&	13.96 	&	11.00 	&	9.62 	&	9.24 	&	8.73 	&	8.83 	&	9.11 \\
				JHR80 D-14	&	12.10 	&	9.95 	&	9.06 	&	8.65 	&	8.48 	&	8.32 	&	8.12 \\
				TIC 190930040	&	13.75 	&	11.16 	&	9.97 	&	9.58 	&	9.25 	&	9.36 	&	9.12 \\
				TIC 345611668	&	12.52 	&	10.72 	&	10.05 	&	9.74 	&	9.64 	&	9.82 	&	9.12 \\
				Elia 3-13	&	8.56 	&	6.53 	&	5.56 	&	5.29 	&	4.71 	&	4.94 	&	4.71 \\
				Elia 3-16	&	10.61 	&	6.97 	&	5.16 	&	4.49 	&	3.33 	&	3.78 	&	3.48 \\
				Elia 3-3	&	8.33 	&	6.59 	&	5.81 	&	5.55 	&	5.32 	&	5.36 	&	5.10 \\
				2MASS J18391505-0605191	&	8.71 	&	6.61 	&	5.62 	&	5.31 	&	4.92 	&	5.06 	&	3.87 \\
				GBM90 Per 75	&	8.88 	&	7.93 	&	7.64 	&	7.44 	&	7.52 	&	7.45 	&	7.39 \\
				TAMURA 8	&	12.53 	&	9.17 	&	7.46 	&	6.83 	&	6.09 	&	6.17 	&	5.91 \\
				TIC 328138536	&	8.61 	&	6.88 	&	6.15 	&	5.86 	&	5.85 	&	5.98 	&	4.83 \\
				IRAS 20315+4026	&	6.96 	&	5.00 	&	4.00 	&	3.56 	&	2.86 	&	3.16 	&	2.81 \\
				TIC 328281708	&	8.71 	&	6.96 	&	6.20 	&	5.71 	&	5.83 	&	5.86 	&	5.03 \\
				TIC 7759115	&	9.37 	&	7.23 	&	6.24 	&	5.64 	&	5.65 	&	5.75 	&	3.63 \\
				WOS 35	&	7.17 	&	5.73 	&	5.05 	&	4.76 	&	4.54 	&	4.68 	&	4.45 \\
				Cl Stephenson 2 9	&	8.43 	&	6.66 	&	5.81 	&	5.32 	&	5.23 	&	5.28 	&	4.17 \\
				TIC 328281041	&	9.12 	&	6.93 	&	5.82 	&	5.15 	&	5.03 	&	4.93 	&	3.17 \\
				StRS 234	&	8.13 	&	6.35 	&	5.51 	&	5.18 	&	4.87 	&	4.82 	&	3.82 \\
				Cl* NGC 7419 BMD 921	&	6.15 	&	4.80 	&	4.23 	&	4.78 	&	3.96 	&	3.82 	&	3.30 \\
				JHR80 D-12	&	11.62 	&	9.53 	&	8.59 	&	8.19 	&	8.13 	&	7.87 	&	7.81 \\
				Cl Stephenson 2 3	&	8.28 	&	6.58 	&	5.80 	&	5.04 	&	5.44 	&	5.27 	&	4.58 \\
				Cl Stephenson 2 8	&	8.24 	&	6.44 	&	5.60 	&	5.02 	&	4.88 	&	4.67 	&	3.52 \\
				Cl Stephenson 2 7	&	8.18 	&	6.45 	&	5.63 	&	5.60 	&	5.12 	&	4.82 	&	3.38 \\
				TIC 150058734	&	10.95 	&	8.19 	&	6.91 	&	6.45 	&	6.08 	&	5.94 	&	5.58 \\
				JHR80 D-11	&	8.66 	&	6.85 	&	6.05 	&	5.66 	&	5.49 	&	5.35 	&	4.95 \\

				HD 188209 (below reference stars)	&	5.72 	&	5.83 	&	5.82 	&	5.79 	&	5.71 	&	5.77 	&	5.53 \\
				* 10 Lac	&	5.30 	&	5.44 	&	5.50 	&	5.63 	&	5.43 	&	5.71 	&	5.70 \\
				* tet Ara	&	4.02 	&	3.93 	&	4.04 	&	4.01 	&	3.72 	&	4.01 	&	3.93 \\
				HR 3091	&	5.86 	&	6.00 	&	5.98 	&	6.01 	&	6.03 	&	6.07 	&	5.50 \\
				* h Car	&	4.22 	&	4.13 	&	4.14 	&	4.00 	&	3.59 	&	3.99 	&	3.89 \\
				* iot Aql	&	4.44 	&	4.42 	&	4.48 	&	4.55 	&	4.26 	&	4.59 	&	4.51 \\
				* eps Sgr	&	1.73 	&	1.77 	&	1.77 	&	1.85 	&	1.26 	&	1.76 	&	1.68 \\
				* N Car	&	4.30 	&	4.22 	&	4.22 	&	4.20 	&	3.89 	&	4.22 	&	4.18 \\
				* n Vel	&	4.34 	&	4.17 	&	4.13 	&	4.04 	&	3.74 	&	4.11 	&	4.03 \\
				HR 8441	&	5.46 	&	5.37 	&	5.33 	&	5.27 	&	5.14 	&	5.34 	&	5.26 \\
				* 22 And	&	4.35 	&	4.05 	&	4.09 	&	3.71 	&	3.49 	&	3.76 	&	3.70 \\
				V* S TrA	&	4.95 	&	4.78 	&	4.59 	&	4.62 	&	4.38 	&	4.59 	&	4.55 \\
				HD 39833	&	6.49 	&	6.23 	&	6.15 	&	6.17 	&	6.07 	&	6.17 	&	6.08 \\
				HD 161178	&	4.30 	&	3.77 	&	3.66 	&	3.52 	&	3.25 	&	3.47 	&	3.44 \\
				* 51 Her	&	3.11 	&	2.56 	&	2.36 	&	2.33 	&	1.88 	&	2.27 	&	2.19 \\
				HD 52418	&	5.00 	&	4.42 	&	4.24 	&	4.17 	&	3.98 	&	4.19 	&	4.14 \\
				HD 42701	&	4.46 	&	3.65 	&	3.60 	&	3.54 	&	3.37 	&	3.53 	&	3.44 \\
				HD 53561	&	5.52 	&	4.53 	&	4.56 	&	4.16 	&	3.89 	&	4.03 	&	3.95 \\
				HD 13570	&	4.78 	&	4.20 	&	3.75 	&	3.77 	&	3.57 	&	3.79 	&	3.68 \\
				HD 189246	&	4.40 	&	3.56 	&	3.42 	&	3.62 	&	3.37 	&	3.51 	&	3.44 \\
				TYC 3440-13-1	&	8.08 	&	7.41 	&	7.20 	&	7.00 	&	6.97 	&	6.89 	&	6.77 \\
				V* VX Pic	&	4.88 	&	4.19 	&	3.93 	&	3.71 	&	3.54 	&	3.55 	&	3.38 \\
				V* V464 Car	&	4.42 	&	3.41 	&	3.12 	&	3.12 	&	3.03 	&	3.05 	&	2.89 \\

	\end{longtable}
\end{footnotesize}

	
	
	\begin{footnotesize}
		\begin{longtable}{l|c|c|c|c|c|c|c|c|c}
			\caption{The peak wavelength ($\lambdapeak$),
				FWHM , and optical depth
				of the 9.7$\mum$ and 18$\mum$ silicate extinction profiles.}\label{table:FWHM} \\
			\hline
			\hline
			Object & $FWHM(9.7)$ & $err_{FWHM(9.7)}$&  $\lambdapeak(9.7)$
			& $err_{\lambdapeak}$&$\Delta\tau_{9.7}$ &$err_{\Delta\tau_{9.7}}$ & $FWHM(18)$ &  $\lambdapeak(18)$ & $\Delta\tau_{18}$ \\
			& ($\mu$m) &  ($\mu$m) & mag & ($\mu$m) &  ($\mu$m) & mag& & & \\
			\hline
			\endfirsthead
			\caption{The peak wavelengths ($\lambdapeak$), FWHMs ,  optical depth of the 9.7$\mum$ and 18$\mum$ silicate extinction profiles.}\\
			
			\hline
			\hline
			Object & $FWHM(9.7)$ & $err_{FWHM(9.7)}$&  $\lambdapeak(9.7)$
			& $err_{\lambdapeak}$&$\Delta\tau_{9.7}$ &$err_{\Delta\tau_{9.7}}$ & $FWHM(18)$ &  $\lambdapeak(18)$ & $\Delta\tau_{18}$ \\
			& ($\mu$m) &  ($\mu$m) & mag & ($\mu$m) &  ($\mu$m) & mag& & &\\
			\hline
			\endhead
			\hline \endfoot
			
            StRS 354	&	2.56	&	0.02 	&	9.73	&	0.01 	&	0.64 	&	0.02 	&	6.17	&	21.92	&	0.28\\
            V* V1074 Sco	&	4.03	&	0.64 	&	9.87	&	0.16 	&	0.11 	&	0.12 	&		&		&	\\
            IC 1795 109	&	2.41	&	0.31 	&	9.82	&	0.06 	&	0.42 	&	0.03 	&		&		&	\\
            VI CYG 1	&	3.51	&	0.13 	&	9.8	&	0.06 	&	0.40 	&	0.06 	&		&		&	\\
            HD 152424	&	1.19	&	0.16 	&	9.72	&	0.11 	&	0.06 	&	0.13 	&		&		&	\\
            HD 229238	&	2.49	&	0.06 	&	9.87	&	0.08 	&	0.23 	&	0.09 	&		&		&	\\
            HD 14956	&	1.58	&	0.44 	&	9.68	&	0.12 	&	0.07 	&	0.66 	&		&		&	\\
            V* DW Cru	&	2.19	&	0.42 	&	9.73	&	0.10 	&	0.15 	&	0.53 	&		&		&	\\
            VI CYG 2	&	3.62	&	0.08 	&	9.83	&	0.08 	&	0.30 	&	0.07 	&		&		&	\\
            HD 147889	&	5.03	&	0.15 	&	9.93	&	0.06 	&	0.31 	&	0.04 	&		&		&	\\
            VI CYG 12	&	2.26	&	0.19 	&	9.78	&	0.04 	&	0.49 	&	0.30 	&	8.13	&	17.72	&	0.14\\
            HD 157038	&	2.03	&	0.32 	&	9.69	&	0.13 	&	0.12 	&	0.65 	&		&		&	\\
            BD+43 3710	&	1.65	&	0.28 	&	9.73	&	0.10 	&	0.17 	&	0.65 	&		&		&	\\
            HD 147701	&	3.64	&	1.12 	&	9.96	&	0.11 	&	0.18 	&	0.09 	&	5.4	&	21.94	&	0.13\\
            StRS 136	&	2.58	&	0.13 	&	9.73	&	0.01 	&	0.71 	&	0.29 	&		&		&	\\
            StRS 164	&	3.04	&	0.06 	&	9.73	&	0.01 	&	0.89 	&	0.05 	&		&		&	\\
            V* V965 Cen	&	1.65	&	0.40 	&	9.68	&	0.16 	&	0.05 	&	0.61 	&		&		&	\\
            HD 29647	&	2.87	&	0.40 	&	9.72	&	0.11 	&	0.16 	&	0.51 	&		&		&	\\
            * phi Cas	&	1.9	&	1.52 	&	9.72	&	0.13 	&	0.06 	&	0.70 	&	4.71	&	19.63	&	0.01\\
            HD 227575	&	2.12	&	0.09 	&	9.87	&	0.08 	&	0.09 	&	0.08 	&		&		&	\\
            HD 283809	&	3.18	&	0.13 	&	9.83	&	0.06 	&	0.30 	&	0.09 	&	9.34	&	21.57	&	0.08\\
            TIC 415790729	&	3.61	&	0.03 	&	9.7	&	0.06 	&	0.87 	&	0.01 	&		&		&	\\
            TIC 268324610	&	4.15	&	0.09 	&	9.72	&	0.01 	&	0.92 	&	0.08 	&		&		&	\\
            Cl* IC 348 LRL 11	&	2.42	&	0.14 	&	9.7	&	0.13 	&	0.33 	&	0.44 	&		&		&	\\
            TIC 345429046	&	4.64	&	0.09 	&	9.77	&	0.00 	&	0.85 	&	0.10 	&		&		&	\\
            JHR80 D-14	&	2.66	&	0.17 	&	9.69	&	0.01 	&	0.59 	&	0.21 	&		&		&	\\
            Elia 3-16	&	4.07	&	0.16 	&	9.71	&	0.03 	&	0.91 	&	0.01 	&		&		&	\\
            TIC 190930040	&	4.74	&	0.06 	&	9.86	&	0.01 	&	0.68 	&	0.02 	&		&		&	\\
            Elia 3-13	&	4.08	&	0.14 	&	9.79	&	0.01 	&	0.54 	&	0.03 	&		&		&	\\
            Elia 3-3	&	5.34	&	0.14 	&	9.68	&	0.06 	&	0.38 	&	0.03 	&		&		&	\\
            TIC 345611668	&	4.6	&	0.21 	&	9.68	&	0.06 	&	0.50 	&	0.04 	&		&		&	\\
            2MASS J18391505-0605191	&	2.55	&	0.07 	&	9.66	&	0.02 	&	0.93 	&	0.21 	&		&		&	\\
            GBM90 Per 75	&	2.33	&	0.29 	&	9.95	&	0.12 	&	0.16 	&	0.80 	&		&		&	\\
            Tamura 8	&	3.78	&	0.10 	&	9.77	&	0.02 	&	0.96 	&	0.13 	&		&		&	\\
            TIC 328138536	&	2.55	&	0.09 	&	9.67	&	0.01 	&	0.68 	&	0.28 	&		&		&	\\
            IRAS 20315+4026	&	2.51	&	0.09 	&	9.73	&	0.00 	&	0.81 	&	0.21 	&	8.88	&	18.44	&	0.27 \\
            TIC 7759115	&	2.55	&	0.08 	&	9.7	&	0.01 	&	0.96 	&	0.19 	&		&		&	\\
            TIC 328281708	&	2.37	&	0.10 	&	9.67	&	0.01 	&	0.66 	&	0.28 	&		&		&	\\
            WOS 35	&	4.05	&	0.15 	&	9.73	&	0.01 	&	0.59 	&	0.29 	&		&		&	\\
            TIC 328281041	&	1.92	&	0.05 	&	9.54	&	0.01 	&	0.74 	&	0.15 	&		&		&	\\
            Cl Stephenson 2 9	&	2.8	&	0.10 	&	9.68	&	0.01 	&	0.73 	&	0.22 	&		&		&	\\
            StRS 234	&	1.82	&	0.07 	&	9.53	&	0.01 	&	0.48 	&	0.23 	&		&		&	\\
            Cl* NGC 7419 BMD 921	&	2.43	&	0.20 	&	9.68	&	0.12 	&	0.32 	&	0.45 	&		&		&	\\
            JHR80 D-12	&	3.81	&	0.17 	&	9.72	&	0.06 	&	0.53 	&	0.18 	&		&		&	\\
            Cl Stephenson 2 8	&	1.63	&	0.06 	&	9.49	&	0.00 	&	0.43 	&	0.23 	&		&		&	\\
            Cl Stephenson 2 3	&	2.17	&	0.10 	&	9.59	&	0.03 	&	0.55 	&	0.23 	&		&		&	\\
            Cl Stephenson 2 7	&	2.15	&	0.07 	&	9.53	&	0.01 	&	0.58 	&	0.20 	&		&		&	\\
            TIC 150058734	&	4.08	&	0.06 	&	9.75	&	0.03 	&	0.63 	&	0.12 	&		&		&	\\
            JHR80 D-11	&	2.74	&	0.19 	&	9.72	&	0.04 	&	0.45 	&	0.33 	&		&		&	\\

		\end{longtable}
	\end{footnotesize}
\end{CJK*}	
\end{document}